\documentclass[10pt,english,aps,prd,superscriptaddress,nofootinbib,notitlepage,preprintnumbers]{revtex4-1}
\usepackage[T1]{fontenc}
\usepackage[utf8]{inputenc}
\usepackage{amsmath}
\usepackage{color}
\usepackage{graphicx}
\graphicspath{{images/}}
\usepackage{wasysym}
\usepackage{esint}
\usepackage[caption=false]{subfig}

\makeatletter

\providecommand{\tabularnewline}{\\}

\def\Mpl{M_{\rm P}}

\usepackage{babel}

\makeatother

\begin{document}
\preprint{YITP-20-64, IPMU 20-0043}
\title{Minimally Modified Gravity fitting Planck data better than $\Lambda$CDM}
\author{Katsuki Aoki}
\email{katsuki.aoki@yukawa.kyoto-u.ac.jp}

\affiliation{Center for Gravitational Physics, Yukawa Institute for Theoretical
Physics, Kyoto University, 606-8502, Kyoto, Japan}
\author{Antonio De~Felice}
\email{antonio.defelice@yukawa.kyoto-u.ac.jp}

\affiliation{Center for Gravitational Physics, Yukawa Institute for Theoretical
Physics, Kyoto University, 606-8502, Kyoto, Japan}
\author{Shinji Mukohyama}
\email{shinji.mukohyama@yukawa.kyoto-u.ac.jp}

\affiliation{Center for Gravitational Physics, Yukawa Institute for Theoretical
Physics, Kyoto University, 606-8502, Kyoto, Japan}
\affiliation{Kavli Institute for the Physics and Mathematics of the Universe (WPI),
The University of Tokyo, Kashiwa, Chiba 277-8583, Japan}
\author{Karim Noui}
\email{Karim.Noui@univ-tours.fr}

\affiliation{APC - Astroparticule et Cosmologie Université Paris Diderot Paris
7, 75013 Paris, France}
\affiliation{Institut Denis Poisson, Université de Tours, Université d'Orléans,
Parc de Grandmont, 37200 Tours, France}
\author{Michele Oliosi}
\email{michele.oliosi@yukawa.kyoto-u.ac.jp}

\affiliation{Center for Gravitational Physics, Yukawa Institute for Theoretical
Physics, Kyoto University, 606-8502, Kyoto, Japan}
\author{Masroor C.~Pookkillath}
\email{masroor.cp@yukawa.kyoto-u.ac.jp}

\affiliation{Center for Gravitational Physics, Yukawa Institute for Theoretical
Physics, Kyoto University, 606-8502, Kyoto, Japan}
\begin{abstract}
We study the phenomenology of a class of minimally modified gravity
theories called $f(\mathcal{H})$ theories, in which the usual general
relativistic Hamiltonian constraint is replaced by a free function
of it. After reviewing the construction of the theory and a consistent
matter coupling, we analyze the dynamics of cosmology at the levels
of both background and perturbations, and present a concrete example
of the theory with a $3$-parameter family of the function $f$. Finally,
we compare this example model to Planck data as well as some later-time
probes, showing that such a realization of $f(\mathcal{H})$ theories
fits the data significantly better than the standard $\Lambda$CDM
model, in particular by modifying gravity at intermediate redshifts,
$z\simeq743$. 
\end{abstract}
\maketitle

\section{Introduction}

The $\Lambda\text{CDM}$ model for cosmology depends on 6 parameters~\cite{Kosowsky:2002zt,Durrer:2008eom}.
Various features of our universe are successfully described by the
$\Lambda\text{CDM}$ model. Setting aside the cosmological constant
problem and other theoretical issues, it provides us the simplest
description of the universe with the cosmic microwave background radiation,
the large-scale structure and the accelerated expansion, the last
of which we probe for example by supernovae observations. Up to now,
this model is still considered the best fitting model to the cosmological
data sets.

Despite the success of the $\Lambda\text{CDM}$ model to describe our
universe, it faces tensions in explaining a few parameters from
various cosmological data sets. This situation motivates us to
consider models beyond $\Lambda\text{CDM}$. The most famous and
significant tension lies between the estimation of today's Hubble
expansion parameter $H_{0}$ from Planck data~\cite{Bernal:2016gxb},
and its measurement from observations of the local universe, such as
those from the Hubble Space Telescope (HST)~\cite{Riess:2019cxk},
$\text{H\ensuremath{0}LiCOW}$~\cite{Wong:2019kwg}, Megamaser Cosmology
Project (MCP)~\cite{Reid:2008nm}, Carnegie-Chicago Hubble Program
(CCHP) Collaboration~\cite{Freedman:2019jwv}, etc.  The significance
of this first tension goes up to $5.3\sigma$~\cite{Wong:2019kwg}.
This discrepancy can be understood as a tension between late time
data, which do not assume any prior cosmological model, and early
universe data, which assume a prior model, i.e. the
$\Lambda\text{CDM}$.  It comes to light as the early universe and the
late time observations (HST) have achieved greater precision in their
measurements. At present, there are several approaches to resolve this
particular tension, including modified gravity (see
~\cite{DeFelice:2020sdq, Ballardini:2020iws, Braglia:2020iik} for
example), or adding new types of matter content (as early dark energy
\cite{Poulin2019} for example).

The number two tension lies between the estimation of the growth of
structure $S_{8}$ from Planck data and that from the redshift space
distortion data. The disagreement has a statistical significance of
up to $3.2\sigma$~\cite{Asgari:2019fkq}. There are several investigations
to address this issue from the perspective of modified gravity theories~\cite{DeFelice:2016ufg,DeFelice:2020icf}.

Two other results could fit the bill as tensions within the $\Lambda$CDM
model. The Planck 2018 result has reported a preference for a higher
lensing amplitude~\cite{Motloch:2018pjy,Motloch:2019gux}. This anomaly
has been argued to be related to a new possible tension in the $\Lambda\text{CDM}$
model of cosmology, known as $\Omega_{K}$ tension, which is investigated
in~\cite{Handley:2019tkm,DiValentino:2019qzk,DiValentino:2020hov}.
In fact the Planck collaboration also has reported that the Planck
data alone prefers a closed universe~\cite{Aghanim:2018eyx}. In
\cite{DiValentino:2020hov}, it has been found that the curvature
parameter of the universe is in tension with the Planck+late time
data and BAO data sets. In their investigation they have found better
fit to closed universe than $\Lambda\text{CDM}$.

The cosmological tensions arising within $\Lambda\text{CDM}$ can
be interpreted as the indication that $\Lambda\text{CDM}$ may be
only a first approximation to a more accurate theory of cosmology.
In this context it is worth investigating the cosmology of modified
theories of gravity to address these various tensions. As mentioned,
there have been several attempts which rely on modified theories of
gravity. In this work, we focus on a certain class of modified theories
of gravity in which there are only two local gravitational degrees
of freedom propagating, dubbed minimally modified gravity (MMG) 
theories in \cite{Lin:2017oow} and then further developed in 
\cite{Aoki:2018zcv,Aoki:2018brq,Mukohyama:2019unx,DeFelice:2020eju} 
(see \cite{DeFelice:2015hla,DeFelice:2015moy,Bolis:2018vzs,DeFelice:2018vza} 
for an earlier example with two local gravitational degrees of freedom 
in the context of massive gravity and also 
\cite{Afshordi:2006ad,Iyonaga:2018vnu,Feng:2019dwu,Gao:2019twq,Aoki:2020lig} 
for more examples of MMG theories). These theories break four dimensional 
diff-invariance keeping the three dimensional diff-invariance
so that the standard framework of cosmological studies can be applied.
Even though breaking four dimensional diff-invariance leads, in general,
to new degrees of freedom in addition to the usual two gravitational
modes, MMG theories have Hamiltonians linear in the lapse function
so that they come with a Hamiltonian constraint that will reduce the
dimension of the physical phase space and leave only two tensorial
modes in the theory.

Recently there has been a Hamiltonian construction for a class of
MMG theories, dubbed $f(\mathcal{H})$ theories~\cite{Mukohyama:2019unx},
in which the standard Hamiltonian constraint $\mathcal{H}$ of general
relativity (GR) is modified to a free function $f(\mathcal{H})$.
This leads to changing the kinetic structure of the theory from GR
in a background-dependent way and provides an opportunity to address
the cosmological tensions.

In this article we study the cosmology in $f(\mathcal{H})$ theories,
considering a model that we have named ``kink'' model as it has
a kink in the first derivative of the free function $f'\left(\mathcal{H}\right)$.
We introduce a consistent coupling to matter which relies on a gauge
fixing in the Hamiltonian as it was done in \cite{Aoki:2018zcv,Carballo-Rubio:2018czn},
and we study linear perturbations so that we can exhibit the no ghost
conditions. We then implement the scalar linear perturbation equations
in the Boltzmann code called CLASS~\cite{Blas:2011rf}, with covariantly
corrected baryon equations of motion~\cite{Pookkillath:2019nkn}.
Subsequently a Monte Carlo sampling, using MontePython~\cite{Brinckmann:2018cvx,Audren:2012wb},
is done against various cosmological data sets. We  consider data
of Planck 2018 with \texttt{planck\_highl\_TTTEEE}, \texttt{planck\_lowl\_EE},
\texttt{planck\_lowl\_TT} polarization~\cite{Aghanim:2019ame,Akrami:2019izv,Aghanim:2018eyx,Akrami:2018odb,Aghanim:2018oex},
of HST observations~\cite{Riess:2019cxk} consisting in the single
data point of the Hubble constant $H_{0}=74.03_{-1.42}^{+1.42}$,
of the baryon acoustic oscillation (BAO)  from 6dF Galaxy Survey~\cite{Beutler:2011hx}
and the Sloan Digital Sky Survey~\cite{Ross:2014qpa,Alam:2016hwk},
and of the joint light curves (JLA) comprised of 740 type Ia supernovae
\cite{Betoule:2014frx}. We refer to all these data sets as Planck2018+HST+BAO+JLA.
The chains are then analyzed using the well-suited GetDist package~\cite{Lewis:2019xzd}. We find that there is a remarkable
improvement with respect to $\Lambda\text{CDM}$ in the likelihood-parameter
$\chi^{2}$ with a difference of $\Delta\chi^{2}=16.6$ for the chosen
data sets. Although there is only a minimal improvement for the $H_{0}$
tension, the other tensions that have appeared within $\Lambda$CDM
have a chance to be addressed by the $f(\mathcal{H})$ theory.

This paper is organized as follows. In section \ref{sec:mmg}, we
review $f(\mathcal{H})$ theories from their inception, and discuss
the particular matter coupling that has to be adopted to avoid the
propagation of unwanted degrees of freedom. In section \ref{sec:bg_pert},
we use the Lagrangian introduced in section \ref{sec:mmg} to derive
the dynamics of the cosmological background, as well as its perturbations.
In section \ref{sec:concrete_models}, we propose a concrete model
where the function $f$ depends on three free parameters and has a
kink in its first derivative. As we will show subsequently, such a
model has a very good fit to cosmological data. Then, in section \ref{sec:results}
we show that the kink model consistently gives a better fitness parameter
than $\Lambda$CDM, considering both early and late time cosmological data sets together. This
comparison is the main result of this paper. Finally, we conclude
in section \ref{sec:conclusion} with a summary of results and a discussion.

\section{Minimally Modified Gravity and $f(H)$ theories}

\label{sec:mmg}

Minimally modified gravity (MMG) theories are modifications of four-dimensional
GR with two local gravitational degrees of freedom. A systematic construction
of gravitational theories with only (up to) two degrees of freedom
has been initiated in~\cite{Lin:2017oow,Aoki:2018zcv,Aoki:2018brq} (see also~\cite{Carballo-Rubio:2018czn}
for a different perspective). The idea consists in renouncing the
invariance under four dimensional diffeomorphisms but keeping the
three dimensional (spatial) diff-invariance. As generically Lorentz-breaking
gravity theories have more than two degrees of freedom, one has to
find the conditions for the theory to possess enough constraints that
would kill the extra degrees of freedom, which would leave us with
(at most) two gravitational modes only. This section is devoted to
review these conditions following~\cite{Mukohyama:2019unx}.

In the first part, we will quickly recall the basis of the construction
of MMG theories as it was done in~\cite{Mukohyama:2019unx} where
the Hamiltonian point of view was adopted. Then, we will show how
$f(\mathcal{H})$ theories naturally emerge as simple but interesting
examples of MMG theories. Finally, we will explain how to couple matter
to these theories without introducing new degrees of freedom in the
gravitational sector, following the construction developed in~\cite{Aoki:2018zcv,Aoki:2018brq}.

\subsection{Modifying the phase space}

The construction of the $f({\cal H})$ class of MMG theories presented
in~\cite{Mukohyama:2019unx} relies on the Hamiltonian formulation
of GR. The idea consists in modifying the phase space of GR, and not
directly the Lagrangian, in such a way that the modified theory remains
invariant under spatial diffeomorphisms only but still propagates
two tensorial degrees of freedom.

Hence, we start with the Arnowitt-Deser-Misner (ADM) parametrization
of the metric, 
\begin{eqnarray}
ds^{2}\;=\;-N^{2}dt^{2}+\gamma_{ij}(dx^{i}+N^{i}dt)(dx^{j}+N^{j}dt)\,,\label{eq:metric_ADM}
\end{eqnarray}
in terms of the lapse function $N$, the shift vector $N^{i}$ and
the induced spatial metric $\gamma_{ij}$. Hereafter, we will denote
by $D_{i}$ the covariant derivative compatible with $\gamma_{ij}$,
we will lower and raise spatial indices by the metric $\gamma_{ij}$
and its inverse $\gamma^{ij}$, and we will use the notation $\gamma$
for the determinant of the spatial metric.

The phase space is parametrized by the usual ten pairs of conjugate
variables, 
\begin{eqnarray}
 &  & \{\gamma_{ij}(x),\pi^{kl}(y)\}=\frac{1}{2}(\delta_{i}^{k}\delta_{j}^{l}+\delta_{i}^{l}\delta_{j}^{k})\,\delta(x-y)\,,\nonumber \\
 &  & \{N^{i}(x),\pi_{j}(y)\}=\delta_{j}^{i}\,\delta(x-y)\,,\label{10pairs}\\
 &  & \{N(x),\pi_{N}(y)\}=\delta(x-y)\,,\nonumber 
\end{eqnarray}
where $\pi^{ij}$, $\pi_{i}$ and $\pi_{N}$ are momenta, and $\delta(x-y)$
is the 3-dimensional $\delta$-distribution.

In GR, $N$ and $N^{i}$ are Lagrange multipliers, thus their conjugate
momenta $\pi_{N}$ and $\pi_{i}$ vanish, which produces a set of
four primary constraints, 
\begin{eqnarray}
\pi_{N}\,\approx\,0\,,\qquad\pi_{i}\,\approx\,0\,.\label{primaryconstraints}
\end{eqnarray}
We are using the notation $\approx$ for the weak equality in the
phase space, i.e. the equality up to constraints. Concerning the momenta
conjugate to $\gamma_{ij}$, they are easily related to the extrinsic
curvature tensor by, 
\begin{eqnarray}
\pi^{ij}=\sqrt{\gamma}\left(K^{ij}-K\gamma^{ij}\right)\,,\qquad K_{ij}\equiv\frac{1}{2N}\left(\dot{\gamma}_{ij}-D_{i}N_{j}-D_{j}N_{i}\right)\,,
\end{eqnarray}
with $K\equiv K^{i}{}_{i}$ being the trace of the extrinsic curvature.
As a consequence, one can immediately compute the Hamiltonian of GR
which takes the very well-known form, 
\begin{eqnarray}
H\;=\;\int d^{3}x\sqrt{\gamma}\left(N{\cal H}_{0}+N^{i}{\cal H}_{i}\right)\,,
\end{eqnarray}
where the Hamiltonian constraint ${\cal H}_{0}$ and the vectorial
(momentum) constraints ${\cal H}_{i}$ are given by, 
\begin{eqnarray}
{\cal H}_{0}\equiv\frac{1}{\gamma}\left(\pi_{ij}\pi^{ij}-\frac{1}{2}\pi^{2}\right)-R\,,\qquad{\cal H}_{i}\equiv-2D^{j}\left(\frac{\pi_{ij}}{\sqrt{\gamma}}\right)\,,\label{GRconstraints}
\end{eqnarray}
with $R$ being the spatial curvature of the metric $\gamma_{ij}$.
The conservation under time evolution of the primary constraints~(\ref{primaryconstraints})
leads to the secondary constraints ${\cal H}_{0}\approx0$ and ${\cal H}_{i}\approx0$,
which form together with (\ref{primaryconstraints}) a set of first
class constraints. The Hamiltonian analysis closes here: as we started
with 10 pairs of variables (\ref{10pairs}) and we found 8 first class
constraints, we end up with the expected 2 tensorial degrees of freedom
of GR.

\medskip{}

In~\cite{Mukohyama:2019unx}, one proposed a deformation of the phase
space of GR requiring that the modified theory remains invariant under
spatial diffeomorphisms only yet still propagates two tensorial degrees
of freedom. In this approach, one starts with the same (non-physical)
phase space parametrized by the usual ten pairs of conjugate variables
(\ref{10pairs}) and one looks for a total Hamiltonian of the form,
\begin{eqnarray}
H_{{\rm def}}\;=\;\int d^{3}x\sqrt{\gamma}\left[{\cal S}(\gamma_{ij},\pi^{ij},R_{ij},N,D_{i})+N^{i}{\cal H}_{i}\right]\,,\label{Hdef}
\end{eqnarray}
where ${\cal S}$ is a three-dimensional scalar constructed from the
variables $(\gamma_{ij},\pi^{ij},R_{ij},N)$ and their spatial derivatives.
It was implicitly assumed that neither $N$ nor $N^{i}$ are dynamical
variables. Requiring that the theory propagates only two degrees of
freedom (or less) leads necessarily to the condition that ${\cal S}$
is an affine function of $N$, i.e. 
\begin{eqnarray}
{\cal S}(\gamma_{ij},\pi^{ij},R_{ij},N,D_{i})\;=\;{\cal V}(\gamma_{ij},\pi^{ij},R_{ij},D_{i})+N\,{\cal H}_{0{\rm def}}(\gamma_{ij},\pi^{ij},R_{ij},D_{i})\,,\label{VandH0def}
\end{eqnarray}
which is, of course, compatible with the fact that $N$ is not dynamical.
Hence, ${\cal H}_{0{\rm def}}$ can be viewed as a deformation of
the usual Hamiltonian constraint of general relativity whereas ${\cal V}$
is a new term. However, the functions ${\cal V}$ and ${\cal H}_{0{\rm def}}$
are not arbitrary and must satisfy extra conditions to ensure that
no degrees of freedom other than the two tensor modes propagate~\cite{Mukohyama:2019unx}.
Even though these conditions have not been solved in full generality
and rigor, it was shown that any theory whose Hamiltonian is given
by (\ref{Hdef}) with 
\begin{eqnarray}
\{{\cal H}_{0{\rm def}}(x)\,,\,{\cal H}_{0{\rm def}}(y)\}\,\approx\,0\,,\label{PoissonH0def}
\end{eqnarray}
while ${\cal V}$ is totally free, defines a MMG theory. In other
words, (\ref{PoissonH0def}) is a sufficient condition for the theory
defined by the Hamiltonian (\ref{Hdef}) to propagate (at most) two
degrees of freedom.

\subsection{Deforming the Hamiltonian constraint: $f(\mathcal{H})$ theories}

Finding all the functions ${\cal H}_{0{\rm def}}(\gamma_{ij},\pi^{ij},R_{ij},N,D_{i})$
which satisfy (\ref{PoissonH0def}) seems a priori to be a highly
complicated problem. The reason is that, for the theory to propagate
gravitational waves, ${\cal H}_{0{\rm def}}$ must depend on both
$\pi^{ij}$ (which contains time derivative of $\gamma_{ij}$) and
three-dimensional curvature terms (which contain gradients of $\gamma_{ij}$),
and the Poisson bracket between two such terms has, in general, a
very complex expression. Fortunately, this problem admits a solution
that we know very well, that is the Hamiltonian constraint ${\cal H}_{0}$
of GR (\ref{GRconstraints}) which satisfies, 
\begin{eqnarray}
\{{\cal H}_{0}[N_{1}]\,,\,{\cal H}_{0}[N_{2}]\}\;=\;{\cal H}_{i}[N_{1}D^{i}N_{2}-N_{2}D^{i}N_{1}]\,\approx\,0\,,\label{PoissonH0H0}
\end{eqnarray}
where ${\cal H}_{0}[N]$ and ${\cal H}_{i}[N^{i}]$ are smeared constraints
defined, for any function $N$ and any vector field $N^{i}$, by the
integrals, 
\begin{eqnarray}
{\cal H}_{0}[N]\,\equiv\,\int d^{3}x\sqrt{\gamma}\,N(x)\,{\cal H}_{0}(x)\,,\qquad{\cal H}_{i}[N^{i}]\,\equiv\,\int d^{3}x\sqrt{\gamma}\,N^{i}(x)\,{\cal H}_{i}(x)\,.
\end{eqnarray}
As it is very well-known, the property (\ref{PoissonH0H0}) is intimately
linked to the invariance under diffeomorphisms of GR. An immediate
consequence is that any deformed Hamiltonian constraint of the form
\begin{eqnarray}
{\cal H}_{0{\rm def}}(x)\;=\;f({\cal H}_{0})\,,\label{fofHdef}
\end{eqnarray}
where $f$ is an arbitrary function, is also a solution of (\ref{PoissonH0def}),
because it satisfies, 
\begin{eqnarray}
\{{\cal H}_{0{\rm def}}[N_{1}]\,,\,{\cal H}_{0{\rm def}}[N_{2}]\}\;=\;{\cal H}_{i}[(f'({\cal H}_{0}))^{2}\left(N_{1}D^{i}N_{2}-N_{2}D^{i}N_{1}\right)]\,\approx\,0\,,\label{PoissonH0defH0def}
\end{eqnarray}
for any $N_{1}$ and $N_{2}$. The MMG theories defined by (\ref{fofHdef})
have been dubbed $f(\mathcal{H})$ theories with reference to $f(R)$
theories. However, contrary to $f(R)$ theories, $f(\mathcal{H})$
theories do not propagate a scalar mode in addition to the tensor
modes. More precisely, $f(\mathcal{H})$ theories were defined with
the additional condition that ${\cal V}=0$ in the Hamiltonian (\ref{VandH0def}),
which we will also assume hereafter.

\medskip{}

By a Legendre transformation, one can easily compute the corresponding
action. Indeed, the equation of motion for $\gamma_{ij}$, 
\begin{eqnarray}
\dot{\gamma}_{ij}\;=\;D_{i}N_{j}+D_{j}N_{i}+\frac{N}{\sqrt{\gamma}}(2\pi_{ij}-\pi\gamma_{ij})f'({\cal H}_{0})\,,
\end{eqnarray}
enables us to relate the momenta $\pi_{ij}$ to the extrinsic curvature
$K_{ij}$ as follows, 
\begin{eqnarray}
K_{ij}\;=\;\frac{f'({\cal H}_{0})}{\sqrt{\gamma}}\left(\pi_{ij}-\frac{1}{2}\pi h_{ij}\right)\,,
\end{eqnarray}
from which we can implicitly obtain $\pi_{ij}$ in terms of $K_{ij}$
because, in general, this equation is non-linear in $\pi_{ij}$. Nonetheless,
one can compute the action which, after a simple calculation, is given
by 
\begin{eqnarray}
S_{\text{grav}}[\gamma_{ij},N,N^{i}]\;=\;\int d^{4}xN\sqrt{\gamma}\left[\frac{2}{f'(C)}(K_{ij}K^{ij}-K^{2})-f(C)\right]\,,\label{minimalLag}
\end{eqnarray}
where $C$ is formally obtained by solving the equation 
\begin{eqnarray}
C\;=\;\frac{K_{ij}K^{ij}-K^{2}}{[f'(C)]^{2}}-R\,.\label{constraintC}
\end{eqnarray}
Hence, we obtained the gravitational part of the action of $f({\cal H})$
theories. Now we are going to explain how to couple consistently this
action to matter.

\subsection{Coupling to matter: problem and solution}

As was shown first in~\cite{Aoki2018,Carballo-Rubio:2018czn}, one
cannot couple matter minimally at this stage of the construction (except
in the trivial case $f(\mathcal{H}_{0})=\mathcal{H}_{0}$) precisely
due to the deformation of the Hamiltonian constraint. As long as there
is no matter, the deformed Hamiltonian constraint $f({\cal H}_{0})$
satisfies the deformed diffeomorphisms algebra \eqref{PoissonH0defH0def}
which ensures that $f({\cal H}_{0})$ remains first class. However,
if one directly adds the matter part to the gravitational constraints
as it is done in GR, one should consider the gravity+matter constraints
\begin{equation}
\bar{\mathcal{H}}_{0\text{def}}=f(\mathcal{H}_{0})+\mathcal{H}_{0\text{,mat}}\,,\qquad\bar{\mathcal{H}}_{i}=\mathcal{H}_{i}+\mathcal{H}_{i\text{,mat}}\,,
\end{equation}
and one shows that the new deformed Hamiltonian constraint $\bar{\mathcal{H}}_{0\text{def}}$
is, in general, no longer first class because 
\begin{align}
\{\bar{\mathcal{H}}_{0\text{def}}[N_{1}],\bar{\mathcal{H}}_{0\text{def}}[N_{2}]\} & ={\cal H}_{i}[(f'({\cal H}_{0}))^{2}\left(N_{1}D^{i}N_{2}-N_{2}D^{i}N_{1}\right)]+\{\mathcal{H}_{0\text{,mat}}[N_{1}],\mathcal{H}_{0\text{,mat}}[N_{2}]\}\nonumber \\
 & \hspace{1em}+\int d^{3}x\,\sqrt{\gamma}f'(\mathcal{H}_{0})\frac{\partial\mathcal{H}_{0}}{\partial\pi^{ij}}\left(N_{1}\frac{\delta\mathcal{H}_{0\text{,mat}}[N_{2}]}{\delta\gamma_{ij}}-N_{2}\frac{\delta\mathcal{H}_{0\text{,mat}}[N_{1}]}{\delta\gamma_{ij}}\right)\nonumber \\
 & \neq\bar{{\cal H}}_{i}[\ldots]\approx0\,.
\end{align}
In short,
$\{\bar{\mathcal{H}}_{0\text{def}}[N_{1}],\bar{\mathcal{H}}_{0\text{def}}[N_{2}]\}$
is not weakly vanishing and then the Hamiltonian constraint is
downgraded to a second class constraint, which leads to the existence
of an extra degree of freedom in the theory. A solution to this
problem was suggested in~\cite{Aoki2018}: it consists in adding a
gauge fixing term at the level of the vacuum Hamiltonian, that will
split the Hamiltonian constraint into a pair of second class
constraints, which then allows to couple matter minimally. We will
follow this procedure, and give an explicit example for a gauge-fixing
in the next subsection.\footnote{One might hope to find a novel matter
  coupling of MMG theories in which the Hamiltonian constraint remains
  first class~\cite{Lin2019}. However, as far as the authors know, all
  such examples considered so far are equivalent to GR with the
  standard matter coupling up to redefinition of Lagrange multipliers
  since all constraints in those examples are equivalent to those of
  GR with the standard matter coupling. A revised version of
  \cite{Lin2019} also discusses the Einstein-frame description of the
  $f(\mathcal{H})$ theory, while in the present paper we adopt the
  Jordan-frame description as the latter is more convenient for a
  multi-component matter system.}

\subsection{Consistent gauge-fixing in the Hamiltonian}

In this subsection, we specify a gauge fixing procedure that can be
used to couple matter fields consistently. By adding the gauge fixing
term thanks to the Lagrange multiplier $\tilde{\lambda}^{i}$, 
\begin{equation}
H_{\text{gf}}=\int d^{3}x\sqrt{\gamma}\,\tilde{\lambda}^{i}\partial_{i}\left(\frac{\pi}{\sqrt{\gamma}}\right)\,,
\end{equation}
to the Hamiltonian of the $f(\mathcal{H})$ theory, we obtain a new
form of the gravitational Hamiltonian given by, 
\begin{equation}
H_{\text{grav}}\equiv H_{\text{def}}+H_{\text{gf}}=\int d^{3}x\sqrt{\gamma}\left[Nf(\mathcal{H}_{0})+\frac{\pi^{ij}}{\sqrt{\gamma}}(D_{i}N_{j}+D_{j}N_{i})-\frac{\pi}{\sqrt{\gamma}}D_{k}\tilde{\lambda}^{k}\right]\,,
\end{equation}
where we used a simple integration by part in the gauge fixing term.

Now, the Hamiltonian equation of motion for $\gamma_{ij}$ leads to
\begin{equation}
K_{ij}=f'(\mathcal{H}_{0})\left(\frac{\pi_{ij}}{\sqrt{\gamma}}-\frac{1}{2}\frac{\pi}{\sqrt{\gamma}}\gamma_{ij}\right)-\frac{1}{2N}\gamma_{ij}D_{k}\tilde{\lambda}^{k}\,,
\end{equation}
which can be solved with respect to $\pi_{ij}$ as 
\begin{equation}
{\pi_{ij}}=\frac{{\sqrt{\gamma}}}{f'(\mathcal{H}_{0})}(K_{ij}-K\gamma_{ij}-N^{-1}\gamma_{ij}D_{k}\tilde{\lambda}^{k})\,.
\end{equation}
Here $\mathcal{H}_{0}$ is viewed as a function of the velocities
$K_{ij}$ (and no more of the momenta $\pi^{ij}$), and is obtained
from the following implicit equation, 
\begin{equation}
\mathcal{H}_{0}=\frac{1}{[f'(\mathcal{H}_{0})]^{2}}\left[K^{ij}K_{ij}-K^{2}-\frac{2K}{N}D_{k}\tilde{\lambda}^{k}-\frac{3}{2N^{2}}(D_{k}\tilde{\lambda}^{k})^{2}\right]-R\,,\label{constraintCgf}
\end{equation}
that generalizes (\ref{constraintC}) in the presence of the gauge
fixing term.

Performing a Legendre transformation, we immediately obtain the Lagrangian
density 
\begin{eqnarray}
\mathcal{L}_{\text{grav}} & = & \pi^{ij}\dot{\gamma}_{ij}-N\sqrt{\gamma}f(\mathcal{H}_{0})-\pi^{ij}(D_{i}N_{j}+D_{j}N_{i})+\pi D_{k}\tilde{\lambda}^{k}\nonumber \\
 & = & \frac{2N\sqrt{\gamma}}{f'(\mathcal{H}_{0})}\left[(K^{ij}K_{ij}-K^{2})-\frac{2K}{N}D_{k}\tilde{\lambda}^{k}-\frac{3}{2N^{2}}(D_{k}\tilde{\lambda}^{k})^{2}\right]-N\sqrt{\gamma}f(\mathcal{H}_{0})\nonumber \\
 & = & N\sqrt{\gamma}\left[2(\mathcal{H}_{0}+R)f'(\mathcal{H}_{0})-f(\mathcal{H}_{0})\right]\,,\label{Lag1}
\end{eqnarray}
where ${\cal H}_{0}$ is still given by (\ref{constraintCgf}). In
order to implement the relation (\ref{constraintCgf}) directly in  the Lagrangian, we consider, instead of (\ref{Lag1}),
the equivalent Lagrangian density 
\begin{eqnarray}
\tilde{\mathcal{L}}_{\text{grav}} & = & \mathcal{L}_{\text{grav}}+N\sqrt{\gamma}\,\lambda_{1}\left[K^{ij}K_{ij}-K^{2}-\frac{2K}{N}D_{k}\tilde{\lambda}^{k}-\frac{3}{2N^{2}}(D_{k}\tilde{\lambda}^{k})^{2}-(C+R)[f'(C)]^{2}\right]\nonumber \\
 & = & N\sqrt{\gamma}\left\{ (C+R)[2-\lambda_{1}f'(C)]f'(C)-f(C)+\lambda_{1}\left[K^{ij}K_{ij}-K^{2}-\frac{2K}{N}D_{k}\tilde{\lambda}^{k}-\frac{3}{2N^{2}}(D_{k}\tilde{\lambda}^{k})^{2}\right]\right\} \,,\label{Lagtilda}
\end{eqnarray}
where the Lagrange multiplier $\lambda_{1}$ ensures that the new
field $C$ satisfies $C=\mathcal{H}_{0}$. At this stage, we can safely
introduce matter fields minimally coupled to the metric \eqref{eq:metric_ADM}
and then we will use this Lagrangian density in the remainder of the
paper.

\section{Cosmology: Background and Linear Perturbations}

\label{sec:bg_pert}

This section aims at studying cosmological properties of $f({\cal H})$
theories whose dynamics is governed by the Lagrangian density~(\ref{Lagtilda})
supplemented with a matter Lagrangian density ${\cal L}_{{\rm mat}}$
of a perfect fluid which is given by the Sorkin-Schutz Lagrangian
density~\cite{Schutz:1970my} (see also \cite{Schutz:1977df,Brown:1992kc,DeFelice:2009bx,Pookkillath:2019nkn}
for example). In the first subsection, we introduce some notations,
compute the equations of motion for a cosmological (homogeneous and
isotropic) background and give preliminary results concerning linear
perturbations about this background. This enables us to extract in
particular linear stability conditions for the tensor modes and the
scalar mode. In the last two subsections, we explain how to implement
the equations for the background and the linear perturbations in the
Boltzmann code solver in order to solve them numerically and to compare
to Planck data and other experimental probes, which we are going to
do in the subsequent sections.

\subsection{Cosmology of $f({\cal H})$ theories: notations and preliminary results}

As we have just announced in introducing this section, we consider
the total Lagrangian density 
\begin{equation}
\mathcal{L}\;=\;\frac{\Mpl^{2}}{2}\,\tilde{\mathcal{L}}_{\text{grav}}+\mathcal{L}_{\mathrm{mat}}\,,\label{totalLag}
\end{equation}
where $\Mpl$ is the Planck mass, $\tilde{\mathcal{L}}_{\text{grav}}$
the $f({\cal H})$ Lagrangian density (\ref{Lagtilda}) and the matter
is supposed to be a perfect fluid whose dynamics is governed by the
Schutz-Sorkin Lagrangian density $\mathcal{L}_{\mathrm{mat}}$, 
\begin{eqnarray}
\mathcal{L}_{{\rm mat}} & = & -N\sqrt{\gamma}[\rho(n)+J^{\mu}\partial_{\mu}\varphi]\,,\qquad n\equiv\sqrt{-J^{\alpha}J^{\beta}g_{\alpha\beta}}\,,\label{defofn}
\end{eqnarray}
where $\varphi$ is a scalar field, $\rho$ is the energy density,
which depends on the number density $n$ defined from the vector $J^{\mu}$
and the metric $g_{\mu\nu}$ according to (\ref{defofn}). A detailed
study of this Lagrangian density can be found in \cite{Pookkillath:2019nkn}.

\medskip{}

To study both the dynamics of the cosmological background and of the
linear perturbations, we consider a three-dimensional spatial metric
given by a perturbed flat Friedmann-Lemaître-Robertson-Walker (FLRW)
geometry, i.e.~in the form 
\begin{equation}
\gamma_{ij}=a(t)^{2}\,(1+2\zeta+h_{ij})+2\,\partial_{i}\partial_{j}E+\frac{1}{2}a(t)\,(\partial_{i}\beta_{j}+ \partial_j \beta_i)\,,\label{Ezeta}
\end{equation}
where $a(t)$ is the scale factor, $\zeta$ and $E$ are scalar perturbations,
$\beta_{i}$ is a vector perturbation and $h_{ij}$ a tensor perturbation.
Notice that we have already decomposed the perturbations following
the usual scalar-vector-tensor decomposition which implies that $\beta_{i}$
is divergenceless whereas $h_{ij}$ is transverse and traceless on
the FLRW background, i.e., 
\begin{equation}
\delta^{ij}\partial_{i}\beta_{j}=0\,,\qquad\delta^{ik}\partial_{i}h_{kj}=0\,,\qquad\delta^{ij}h_{ij}=0\,.
\end{equation}
The remaining components of the metric, the lapse function and shift
vector, are parametrized by 
\begin{eqnarray}
N=N(t)\,(1+\alpha)\,,\qquad N_{i}=N(t)(\,\partial_{i}\chi+\chi_{i})\,,\label{alphachi}
\end{eqnarray}
where $\alpha$ and $\chi$ are scalar perturbations whereas $\chi_{i}$
is a divergenceless vectorial perturbation, i.e.~$\delta^{ij}\partial_{i}\chi_{j}=0$.
To finish with the variables entering in the gravitational Lagrangian
density, one should also perturb the auxiliary fields $\lambda_{1}$,
$\tilde{\lambda}_{i}$ and $C$ according to, 
\begin{eqnarray}
\lambda_{1}=\lambda_{1}(t)+\delta\lambda_{1}\,,\qquad\tilde{\lambda}^{i}=\frac{1}{a(t)^{2}}\delta^{ij}(\delta\lambda_{2j}+\partial_{j}\delta\lambda_{2})\,,\qquad C=C(t)+\delta C\,,\label{lambdaC}
\end{eqnarray}
where, again, $\delta\lambda_{1}$, $\delta\lambda_{2}$, $\delta C$
are scalar perturbations, and $\delta\lambda_{2i}$ is a divergenceless
vectorial perturbation, i.e. $\delta^{ij}\partial_{i}\delta\lambda_{2j}=0$.
Note that, as usual, the background quantities are time-dependent
only, and the values of $\tilde{\lambda}^{i}$ and $N_{i}$ vanish
due to their vectorial nature and the background symmetry.

Concerning the variables entering specifically in the matter Lagrangian
density (\ref{defofn}), they are parametrized as follows 
\begin{eqnarray}
J^{0}=J^{0}(t)\,(1+\delta J)\,,\qquad J^{i}=\frac{1}{a^{2}}\,\delta^{il}\,\partial_{l}\delta j\,,\qquad\varphi=\varphi(t)-\rho_{,n}\,v\,,\label{eq:pert_dJ}
\end{eqnarray}
where $\delta J$, $\delta j$, and $v$ are scalar perturbations.

\medskip{}

Now, we have introduced all the necessary variables to study the dynamics
of the background and of the linear perturbations. For that purpose,
we expand the total Lagrangian density (\ref{totalLag}) up to the
second order in perturbations. The Lagrangian density linear in the
perturbations provides the following equations of motion for the background,
\begin{eqnarray}
 &  & J^{0}(t)=\frac{\mathcal{N}_{{\rm tot}}}{N\,a^{3}}\,\qquad\varphi(t)=-\int^{t}N\,\rho_{,n}dt'\,,\qquad\lambda_{1}(t)=\frac{1}{f_{,C}}\,,\qquad\frac{\Mpl^{2}}{2}\,f+\rho=0\,,\label{Fried1}\\
 &  & C(t)=-\frac{6}{f_{,C}^{2}}\frac{\dot{a}^{2}}{N^{2}a^{2}}=-\frac{6}{f_{,C}^{2}}\,H^{2}\,,\qquad\frac{f_{,CC}}{f_{,C}^{2}}=\frac{2\Mpl^{2}\dot{H}/N+f_{,C}(\rho+P)}{12H^{2}(\rho+P)}\,,\qquad\frac{\dot{\rho}}{N}=-3\,H\,(\rho+P)\,,\label{eq:conservation-eq}
\end{eqnarray}
where $f_{,C}$ and $f_{,CC}$ are respectively the first and second
derivatives of $f$ with respect to $C$, $H=\dot{a}/(aN)$ is the
Hubble constant and the so-called total particle number ${\cal N}_{{\rm tot}}$
is an integration constant obtained from integrating the conservation
constraint $\nabla_{\mu}J^{\mu}=0$ on a flat FLRW geometry. The last
equation of \eqref{Fried1} and the last two equations of \eqref{eq:conservation-eq}
are the modified Friedmann equations and the usual conservation equation
for the minimally coupled matter. It is easy to verify that, in the
limit $f_{,C}\to1$, we recover the GR equations of motion coupled
to a perfect fluid. Taking the time derivative of the first equation
of \eqref{eq:conservation-eq}, and using the modified Friedmann equations
together with the conservation equation, we find the useful relation,
\begin{equation}
0=\frac{\Mpl^{2}}{2}\,f_{,C}\,\dot{C}-3\,N\,H\,(\rho+P)=0\quad\Longleftrightarrow\quad\frac{\dot{C}}{N}=\frac{6\,H\,(\rho+P)}{\Mpl^{2}\,f_{,C}}\,.
\end{equation}
Therefore, the equations of motion enable us to express the four quantities
$f,\dot{H}$, $C$ and $\dot{C}$ in terms of $f_{,C}$, $f_{,CC}$
and the other standard cosmological quantities $a,H,\rho,\dots$ only.
Hence, we can eliminate the variables $f,\dot{H}$, $C$ and $\dot{C}$
by a direct substitution, which we are going to do from now on.

Expanding the Lagrangian density at second order in perturbations
and Fourier transforming as usual with respect to the spatial comoving
coordinates, we can then find, for high $k\equiv|\vec{k}|$ (with
$\vec{k}$ the comoving momentum), the no-ghost conditions, together
with the speeds of propagation. Tensor perturbations give 
\begin{eqnarray}
Q_{T}=(f_{,C})^{-1}>0\,,\qquad c_{T}^{2}=(f_{,C})^{2}\,,\label{eq:cT2}
\end{eqnarray}
so that letting $f_{,C}>0$ at all times is sufficient to avoid the
presence of ghost-like tensor modes. Furthermore, the famous constraint
on the speed of propagation of gravitational waves today   \cite{TheLIGOScientific:2017qsa} (at redshift
$z=0$) imposes that $c_{T}\approx1$ to the accuracy of order $\mathcal{O}(10^{-15})$,
which leads to the condition that $(f_{,C})^{2}\approx1$ to the accuracy
of order $\mathcal{O}(10^{-15})$ at $z=0$ as well.

The analysis of scalar perturbations gives, 
\begin{eqnarray}
Q_{s} & = & \frac{\rho^{2}a^{2}}{2k^{2}n\rho_{,n}}=\frac{\rho^{2}a^{2}}{2k^{2}\,(\rho+P)}>0\,,\\
c_{s}^{2} & = & \frac{n\rho_{,nn}}{\rho_{,n}}+\frac{2}{\Mpl^{2}}\,\frac{f_{,CC}}{f_{,C}^{2}}\,(\rho+P)\,.\label{eq:cs2}
\end{eqnarray}
Hence, we find that as long as $f_{,CC}\neq0$, the speed of propagation
changes for any matter component, dust included. Finally, and as expected,
there are no propagating vector perturbations in the gravity sector.

\subsection{Implementation of background equations of motion}

Now, we are going to explain in detail the way we have implemented
the background equations of motion in the code. First, we extend the
theory to the case where several fluids are coupled to gravity (in
order to model radiation, dust and eventually dark energy components) by adding
to the total Lagrangian density \eqref{totalLag} different matter
Lagrangian densities. In that case, the two modified Friedmann equation
and the conservation equation given in~(\ref{Fried1}, \ref{eq:conservation-eq})
become, 
\begin{equation}
\frac{\Mpl^{2}}{2}\,f+\sum_{i}\rho_{i}=0\,,\quad\frac{f_{,CC}}{f_{,C}^{2}}=\frac{2\Mpl^{2}\dot{H}/N+f_{,C}\sum_{i}(\rho_{i}+P_{i})}{12H^{2}\sum_{i}(\rho_{i}+P_{i})}\,,\quad\frac{\dot{\rho}}{N}=-3\,H\,\sum_{i}(\rho_{i}+P_{i})\,,\label{eq:Fried_multi}
\end{equation}
where the subscript $i$ parametrizes the different fluids, while
the remaining equations of motion are unchanged, 
\begin{eqnarray}
J_{i}^{0}(t)=\frac{\mathcal{N}_{i{\rm tot}}}{N\,a^{3}}\,\quad\varphi_{i}(t)=-\int^{t}N\,\rho_{i,n}dt'\,,\quad\lambda_{1}(t)=\frac{1}{f_{,C}}\,,\quad C(t)=-\frac{6}{f_{,C}^{2}}\,H^{2}\,.\label{Ceq}
\end{eqnarray}
On using the equation for $C(t)$ in (\ref{Ceq}), the first modified
Friedmann equation in (\ref{eq:Fried_multi}) can be rewritten as
\begin{equation}
H^{2}=\varrho_{f}+\sum_{i}\varrho_{i}\qquad{\rm or\,\,equivalently}\qquad1=\Omega_{f}+\sum_{i}\Omega_{i}\,,\label{eqforOmega}
\end{equation}
where we used the notations, 
\begin{equation}
\varrho_{f}\equiv\frac{\rho_{f}}{3\Mpl^{2}}\equiv\frac{1}{6}\,(f-C\,f_{,C}^{2})\,,\qquad\varrho_{i}\equiv\frac{\rho_{i}}{3\Mpl^{2}}\,,\qquad\Omega_{f}\equiv\frac{\varrho_{f}}{H^{2}}\,,\qquad\Omega_{i}\equiv\frac{\varrho_{i}}{H^{2}}\,.\label{eq:varrho_f}
\end{equation}
Hence, the modification of GR shows up, in the first Friedmann equation,
as a fluid of effective density $\rho_{f}$. Following a similar procedure
we can also find the expression for the effective pressure $P_{f}$
or equivalently $p_{f}\equiv P_{f}/(3\Mpl^{2})$. Indeed, using the
definitions of $\varrho_{f}$ and $\varrho_{i}$ given in (\ref{eq:varrho_f}),
the second modified Friedmann equation in \eqref{eq:Fried_multi}
can be reformulated as follows, 
\begin{equation}
\frac{2}{3}\frac{\dot{H}}{a}+\left(\varrho_{f}+p_{f}\right)+\sum_{i}(\varrho_{i}+p_{i})=0\,,\label{eq:secondaE}
\end{equation}
where we used the notations, 
\begin{equation}
p_{f}=-(1-f_{,C}-2\,Cf_{,{\it CC}})\sum_{i}(\varrho_{i}+p_{i})-\frac{1}{6}\,(f-C\,f_{,C}^{2})\,,\qquad p_{i}=\frac{P_{i}}{3\Mpl^{2}}\,.
\end{equation}
Hence, $p_{f}$ is the (normalized) pressure of the effective fluid
which describes the modifications of GR in a cosmological background.
Finally, we also define the effective equation of state $w_{f}\equiv p_{f}/\varrho_{f}$.

\medskip{}

Now, we are ready to show how to implement these equations of motion
in the code. We start by choosing the lapse function so that it coincides
with the scale factor, i.e.\ $N=a$, and then the time $t$ becomes
the conformal time $\tau$. Hence, from now on, an overdot represents
the derivative with respect to $\tau$, i.e., 
\begin{eqnarray}
\dot{F}\equiv\frac{dF}{d\tau}\,,
\end{eqnarray}
for any function $F$. This choice is motivated by the fact that most
of the Boltzmann solvers are written in terms of the conformal time.
It is also convenient, from this section on, to introduce the derivative
with respect to $N_{e}\equiv-\ln(a/a_{0})$, where $a_{0}$ is the
scale factor today (at $z=0$) for any function $F$. The $\tau$
and $N_{e}$ derivatives are related by 
\begin{equation}
\dot{F}=\frac{da}{d\tau}\,\,\frac{dF}{da}=\dot{a}\,\frac{dN_{e}}{da}\,\frac{dF}{dN_{e}}=-\frac{\dot{a}}{a}\,\frac{dF}{dN_{e}}=-a\,H\,\frac{dF}{dN_{e}}\,.
\end{equation}

Then, we differentiate the first Friedmann equation in \eqref{eq:Fried_multi}
with respect to $\tau$ and, after using the conservation equations~(\ref{eq:conservation-eq})
for each fluid, we obtain the dynamical equation, 
\begin{equation}
\dot{C}=\frac{6\,a\,H\,\sum_{i}(\rho_{i}+P_{i})}{\Mpl^{2}f_{,C}}=\frac{18\,a\,H\,\sum_{i}(\varrho_{i}+p_{i})}{f_{,C}}\,,\label{eq:dot_C}
\end{equation}
which is equivalent to 
\begin{eqnarray}
\frac{dC}{dN_{e}}=-\frac{18\,\sum_{i}(\varrho_{i}+p_{i})}{f_{,C}}\,.
\end{eqnarray}
As a consequence, the dark energy component does not contribute to
the derivative of $C$ and, if we denote by $\rho_{m}$ the dark matter
density and by $\rho_{r}$ the radiation density, we have 
\begin{eqnarray}
\left\{ \begin{array}{ll}
\dfrac{dC}{dN_{e}} & =-{18\,(\varrho_{m}+4\varrho_{r}/3)}/{f_{,C}}=-{6\,H_{0}^{2}\,(3\Omega_{m0}a+4\,\Omega_{r0})}/({f_{,C}\,a^{4}})\,,\vspace{1mm}\\
\dfrac{da}{dN_{e}} & =-a\,,
\end{array}\right.\label{eq:ODEs}
\end{eqnarray}
where $H_{0}$ and $\Omega_{i0}$ are the values of $H$ and $\Omega_{i}$
today (at $z=0$). The second differential equation has been introduced
so that we obtain an autonomous system of Ordinary Differential Equations
(ODEs). We need to provide the initial conditions, namely $C(N_{e}=0)=C_{0}$
and $a(0)=1$ to fully integrate the system\footnote{As we need to find the background evolution at very high redshifts,
we need a stable integrator. We found it in a Runge-Kutta Cash-Karp
(4, 5) method. Instead for the Boltzmann code involving also the equations
of motion for the perturbations we have used the default integrator
of CLASS.}. Therefore, in order to have $C$ at any value of the redshift $z$
($z>0$) we just need to integrate the system of ODEs up to $N_{e}=\ln(1+z)$.
Finally, on solving the system of ODEs (\ref{eq:ODEs}) for a given
function $f(C)$ we are also able to find at any redshift the value
of every relevant background quantity.

Let us now discuss the initial conditions. From \eqref{eqforOmega},
we know that today $\Omega_{f0}=1-\sum_{i}\Omega_{i0}$. On combining
this relation with the first Friedmann equation in (\ref{eq:Fried_multi}),
we find that 
\begin{eqnarray}
f(z=0)=-6H_{0}^{2}(1-\Omega_{f0})\,,\qquad[f_{,C}(z=0)]^{2}\,C_{0}=-6H_{0}^{2}\,.\label{initialconditions}
\end{eqnarray}
Furthermore, since we impose from observations that $[f_{,C}(z=0)]^{2}\approx1$
to the accuracy of order $\mathcal{O}(10^{-15})$ (see discussion
below Eq. \eqref{eq:cT2}), we find that $C_{0}\approx-6H_{0}^{2}$,
and we can safely\footnote{One could solve the initial condition equation $C_{0}=-6H_{0}^{2}/[f_{,C}(z=0)]^{2}$
for $C_{0}$ iteratively. At lowest order, one finds $C_{0}=-6H_{0}^{2}$,
and at next order $C_{0}=-6H_{0}^{2}/[f_{,C}(C=-6H_{0}^{2})]^{2}$.
However, for the models we have considered, the second solution is
numerically indistinguishable from the value $C_{0}=-6H_{0}^{2}$.} set this value as the initial condition for $C$.

\subsection{Implementation of perturbation equations}

In this subsection, we are going to compute the equations for the
linear scalar perturbations and reduce them to a minimal but complete
system where we have eliminated the redundant variables. We start
with $(7+3\,s)$ scalar perturbations (where $s$ is the number of
matter components): for the metric, we introduced $\zeta$, $E$,
$\alpha$, $\chi$ in \eqref{Ezeta}-\eqref{alphachi}, for the matter
components we introduced $\delta J_{i}$, $\delta j_{i}$ and $v_{i}$
in \eqref{eq:pert_dJ}, and finally we introduced $\delta\lambda_{1}$,
$\delta\lambda_{2}$, $\delta C$ in \eqref{lambdaC} for the remaining
variables. As usual, we need to expand the total Lagrangian density
for $f(\mathcal{H})$ theories coupled to matter fields up to second
order and we perform a Fourier transformation with respect to the
spatial comoving coordinates to obtain the  equations of the perturbations.
In the sequel, we will denote by $E_{Q}=0$ the equation of motion
of the perturbation variable $Q$ obtained by deriving the quadratic
Lagrangian density with respect to $Q$. For instance, we find 
\begin{equation}
\frac{E_{\chi}}{\Mpl^{2}}=-2\,k^{2}a^{2}H\,\delta\lambda_{1}+\frac{k^{4}}{a\,f_{,C}}\,\delta\lambda_{2}+\frac{2k^{2}a^{2}H}{f_{,C}}\,\alpha+3\,k^{2}a^{2}\sum_{i}n_{i}\varrho_{i,n}v_{i}-\frac{2k^{2}a}{f_{,C}}\,\dot{\zeta}=0\,,
\end{equation}
and similar equations can be computed for the other variables. Of
course, we make use of the equations of motion for the background
to simplify several expressions, in particular, the background lapse
function still coincides with the scale factor, namely $N(\tau)=a(\tau)$,
where $\tau$ is the conformal time.

The first step consists in eliminating the redundant variables. We
can see that we can integrate out the auxiliary variables $\delta J_{i}$,
$\delta j_{i}$ in the matter Lagrangian density, only leaving the
fields $v_{i}$ and $\delta_{i}$ for each matter field components
(see \cite{Pookkillath:2019nkn} for details). In the gravitational
sector, we make use of the spatial gauge freedom (the invariance under
three-dimensional diffeomorphisms) to fix $E=0$ as it is often done.
Furthermore, as we are going to see later on, it is convenient to
use the same gauge invariant combinations of fields as the ones used
in $\Lambda$CDM to describe perturbations in the Newtonian gauge.
Hence, we make the following field redefinitions, 
\begin{eqnarray}
\frac{\delta\rho_{i}}{\rho_{i}}\equiv\delta_{i}+\frac{3\dot{a}n_{i}\varrho_{i,n}}{a^{2}\,\varrho_{i}}\,\chi\,,\qquad\alpha\equiv\psi-\frac{\dot{\chi}}{a}\,,\qquad\zeta\equiv-\phi-\frac{\dot{a}}{a^{2}}\,\chi\,,\qquad v_{i}\equiv-\frac{a}{k^{2}}\,\theta_{i}+\chi\,.\label{eq:def-thetai}
\end{eqnarray}
Again, these combinations, in the language of GR, correspond to the
gauge-invariant combinations which reduce to the Newtonian fields
(when we adopt the Newtonian gauge).

\medskip{}

Now, we have all the ingredients to compute and simplify the equations
for the perturbations. We start with the matter equations which are
given by, 
\begin{eqnarray}
E_{v}^{(i)} & \equiv & \dot{\delta}_{i}+\frac{\varrho_{i}+p_{i}}{\varrho_{i}}\,\theta_{i}-\frac{3\dot{a}}{a}\left(\frac{p_{i}}{\varrho_{i}}-\frac{n_{i}\varrho_{i,nn}}{\varrho_{i,n}}\right)\delta_{i}-\frac{3(\varrho_{i}+p_{i})}{\varrho_{i}}\,\dot{\phi}=0\,,\label{Eqmatter1}\\
E_{\delta\rho}^{(i)} & \equiv & \dot{\theta}_{i}-k^{2}\psi-\frac{\varrho_{i}}{\varrho_{i}+p_{i}}\,\frac{n_{i}\varrho_{i,nn}}{\varrho_{i,n}}\,k^{2}\,\delta_{i}+\frac{\dot{a}}{a}\left(1-\frac{3n_{i}\varrho_{i,nn}}{\varrho_{i,n}}\right)\theta_{i}+k^{2}\sigma_{i}=0\,,\label{Eqmatter2}
\end{eqnarray}
where ${n_{i}\varrho_{i,nn}}/{\varrho_{i,n}}=\dot{p}_{i}/\dot{\varrho}_{i}$
and $\sigma_{i}$ is the shear perturbation for each fluid. We see
that the form of the equations of motion in the matter sector is identical
to the ones in GR in the Newtonian gauge. This is a consequence of
the fact that the auxiliary field $C$ in the $f(\mathcal{H})$ theory
does not have any direct coupling with matter fields, and matter is
minimally coupled to the gravitational sector.

\medskip{}

Then, we study the equations $E_{\delta\lambda_{1}}$, $E_{\delta\lambda_{2}}$
and $E_{\delta C}$. After a direct calculation, we see that we can
integrate out the variables $\delta\lambda_{1}$, $\delta C$ and $\delta\lambda_{2}$
according to, 
\begin{eqnarray}
\delta\lambda_{1} & = & -\frac{f_{,CC}}{f_{,C}^{2}}\,\delta C\,,\qquad\delta C=\frac{4k^{2}}{a^{2}}\,(\phi+H\chi)\,,\\
\delta\lambda_{2} & = & \left[{\frac{a^{2}(f_{{,C}}-12\,f_{,CC}H^{2})}{6f_{{,C}}H{k}^{2}}}-\frac{a^{3}\dot{H}}{2H{k}^{4}}\right]a\,\delta C-{\frac{2a^{2}\dot{\phi}}{{k}^{2}}}+\left({\frac{2a\dot{H}}{H{k}^{2}}}-{\frac{2}{3H}}\right)a\,\phi-{\frac{2a^{3}H\psi}{{k}^{2}}}\,.
\end{eqnarray}
Furthermore, on considering the equation of motion $E_{\chi}=0$,
we can also find $\chi$ as follows,
\begin{equation}
\chi=\frac{9af_{,C}}{k^{2}[9f_{,C}\sum_{j}(\varrho_{j}+p_{j})+2\,k^{2}/a^{2}]}\,\sum_{i}(\varrho_{i}+p_{i})\theta_{i}\,.
\end{equation}

\medskip{}

Finally, it remains to study the equations which determine the dynamics
for the perturbations in the gravitational sector, namely for $\phi$
and $\psi$. We start by substituting the Newtonian gauge fields \eqref{eq:def-thetai}
directly in the total Lagrangian density ${\cal L}$ so that we obtain
a new equivalent Lagrangian density $\mathcal{L}'$ for these new
variables. Interestingly, we find that the equation for $\chi$, that
we denote $\bar{E}_{\chi}=0$, is a new one and can be written as
\begin{eqnarray}
E_{1}\equiv\dot{\phi}+aH\,\psi+\frac{4k^{2}Hf_{,CC}}{af_{,C}}\,\phi+\frac{3a^{2}}{\Gamma_{1}k^{2}}\left[6\Gamma_{2}H^{2}\frac{f_{,CC}}{f_{,C}}-\frac{1}{2}\Gamma a^{2}f_{,C}^{2}-\frac{k^{2}\Gamma_{1}}{9\Gamma_{2}}\right]\sum_{i}(\varrho_{i}+p_{i})\,\theta_{i}=0\,,\label{eq:fond_eq_1}
\end{eqnarray}
where we have introduced the notations, 
\begin{equation}
\Gamma\equiv\sum_{i}n_{i}\varrho_{i,n}=\sum_{i}(\varrho_{i}+p_{i})\,,\qquad\Gamma_{1}\equiv\Gamma\,a^{2}f_{,C}+\frac{2}{9}k^{2}\,,\qquad\Gamma_{2}\equiv\Gamma\,a^{2}+\frac{2}{9}k^{2}f_{,C}\,.
\end{equation}
In the limit where $f_{,C}\to1$ and $f_{,CC}\to0$, we recover the
expected results of GR and, (\ref{eq:fond_eq_1}) represents one of
the two fundamental equations which involve the metric perturbation
variables, that Boltzmann solvers need to solve besides the equations
of motion for the matter variables.

At this stage, we need another equation for $\psi$. We find it by
the following procedure. First of all we consider $\bar{E}_{\psi}=0$.
This equation can be written as 
\begin{equation}
\frac{\bar{E}_{\psi}}{3a^{4}H\Mpl^{2}}=-\frac{2}{3}\,f_{,C}\,\frac{k^{2}}{a^{2}H}\,\phi-\frac{1}{H}\sum_{i}\varrho_{i}\,\delta_{i}-\frac{3a\Gamma_{2}}{k^{2}\Gamma_{1}}\sum_{i}n_{i}\varrho_{i,n}\theta_{i}=0\,.
\end{equation}
Then, we compute $\dot{\bar{E}}_{\psi}$ and we obtain an equation
that involves the terms $\dot{\phi}$, $\dot{\delta}_{i}$ and $\dot{\theta}_{i}$.
Therefore from linear combinations of this equation with $E_{1}$
and the matter equations of motion, as well as $\bar{E}_{\psi}$,
we can remove each of these derivatives. Finally we obtain the remaining
expected equation, 
\begin{eqnarray}
E_{2} & \equiv & \psi+\frac{(4f_{,CC}k^{2}-a^{2}f_{,C})\Gamma_{1}}{a^{2}\Gamma_{2}}\,\phi+\frac{9a^{2}f_{,C}}{2k^{2}}\sum_{i}(\varrho_{i}+p_{i})\sigma_{i}-\frac{a^{2}(f_{,C}^{2}-1)}{\Gamma_{2}}\,\sum_{i}\frac{n_{i}\varrho_{i,nn}}{\varrho_{i,n}}\,\varrho_{i}\,\delta_{i}\nonumber \\
 & + & \frac{1}{k^{2}\Gamma_{1}\Gamma_{2}}\left[a^{4}f_{,C}(f_{,C}^{2}-1)(\dot{\Gamma}+2Ha\Gamma)+2Ha\frac{f_{,CC}}{f_{,C}}\left(2f_{,C}^{2}k^{2}\Gamma_{1}-9a^{2}\Gamma\Gamma_{2}\right)\right]\sum_{i}(\varrho_{i}+p_{i})\theta_{i}\,=0\,,\label{eq:fond_eq_2}
\end{eqnarray}
which, in the GR limit, reduces to the correct shear perturbation
equation.

\medskip{}

We have now all the dynamical equations of motion necessary to implement
the dynamics of the whole system: the matter equations \eqref{Eqmatter1}
and \eqref{Eqmatter2} together with the ``gravitational'' equations
\eqref{eq:fond_eq_1} and \eqref{eq:fond_eq_2} couple the perturbation
variables $\theta_{i},\delta_{i},\phi$ and $\psi$. Notice that this
theory does not add any new propagating degree of freedom and this
is the reason why we do not need to add any new equation to the Boltzmann
code. The main difference between this theory and GR lies, at the
level of perturbations, on the two modified equations of motion for
the metric perturbation variables.

\section{A Concrete example: the ``kink'' model}

\label{sec:concrete_models}

In the following, we will consider one explicit example of $f(H)$ theories by choosing a particular class of functions for $f$.
Before presenting the model in details, we give our motivations and
also some of the most important results we have obtained with it.

This model has been built so that it is consistent with the constraint
$c_{T}=1$ at low redshifts, which means $f_{,C}(z=0)\approx1$, and
it possibly addresses some so-far unsolved cosmological puzzles. We
have found that the model we are to introduce is capable of fitting
the chosen data sets better than $\Lambda$CDM. Even if this model
does not improve much the fit to the $H_{0}$ measurement, it achieves
a better fit especially for Planck data. Furthermore, although in
the context of spatially curved ($\Omega_{K0}\ne0$) $\Lambda$CDM,
one is able to fit Planck data alone better, as soon as we introduce
late time data, BAO data in particular, $\Omega_{K0}\neq0$ is strongly
constrained. Some authors \cite{DiValentino:2020hov} are led to state
that $\Lambda$CDM is ruled out in the light of such behavior. However,
the model we are going to introduce, does not feel such a strong constraint
from late time data. Before and after the insertion of late time data,
Planck data are considerably fit better by the model we are about
to discuss. Hence, this study opens a window on using modified gravity
models to address not only late time cosmology but even cosmology
at intermediate/high redshifts.

\medskip{}

Before defining our model, let us remind that $\Lambda$CDM corresponds
to the case where $f(C)$ is an affine function of $C$, i.e. of the
form $f(C)=C-C_{0}+f_{0}$ with  $f_0 \equiv f(C_{0})$. The
model we are now considering consists in choosing $f(C)$ such that
its derivative is of the form, 
\begin{equation}
f_{,C}=1+\frac{1}{2}a_{1}-\frac{1}{2}a_{1}\tanh\!\left[\frac{1}{a_{3}}\left(\frac{C}{H_{0}^{2}}+a_{2}\right)\right],
\end{equation}
where $a_{1}$, $a_{2}$ and $a_{3}$ are free positive real parameters. The fact that $a_1>0$ ensures $f_{,C}>0$ which is the condition  to avoid the presence of ghost-like tensor modes \eqref{eq:cT2}. 
 The hyperbolic tangent
function gives $f_{,C}$ a kink shape and, for this reason, we dub
it the ``kink''-model. A direct integration with respect to the
$C$ variable leads to 
\begin{equation}
f(C)=f_{0}+(1+a_{1})\,(C-C_{0})+\frac{a_{1}a_{3}H_{0}^{2}}{2}\left[{\rm softplus}\!\left(\frac{2a_{2}H_{0}^{2}+2C_{0}}{a_{3}H_{0}^{2}}\right)-{\rm softplus}\!\left(\frac{2a_{2}H_{0}^{2}+2C}{a_{3}H_{0}^{2}}\right)\right],\label{eq:asymmetricbump}
\end{equation}
where $f_{0}$ is the integration constant, $C_{0}$ is the value
of $C$ today, and we have made use of the softplus function\footnote{The softplus function is defined as ${\rm softplus}(x)=\ln(1+e^{x})$.
An equivalent definition, better suited for numerical purposes, is
${\rm softplus}(x)={\rm max}(0,x)+\ln(1+e^{-|x|})$.}.

Some of the free parameters are constrained by the initial conditions
\eqref{initialconditions}. Indeed, as $C_{0}\approx-6H_{0}^{2}$
and $f_{,C}\approx1$ at $z=0$, then the parameters $a_{2}$ and
$a_{3}$ must satisfy the condition $(a_{2}-6)/a_{3}\gg1$ which implies
that $a_{2}\gg a_{3}$. Furthermore, on using the Friedmann equation,
we already know that $f_{0}=-6H_{0}^{2}(1-\Omega_{f0})$.

Now, let us study some properties of the model at very early times
when $C<0$ and $|C|\gg a_{2}H_{0}^{2}$. In that case, the function
$f(C)$ simplifies and becomes, 
\begin{equation}
f(C)\approx(1+a_{1})\,(C+6H_{0}^{2})+f_{0,{\rm eff}}\,\,,\qquad f_{0,{\rm eff}}\equiv6H_{0}^{2}\Omega_{f0}-6H_{0}^{2}+\frac{a_{1}a_{3}H_{0}^{2}}{2}\,{\rm softplus}\!\left(\frac{2a_{2}-12}{a_{3}}\right).
\end{equation}
As a consequence, at early times, the theory is equivalent to a GR
 theory with $f(C)$ an affine function of $C$, i.e. of the form
$f(C)=\alpha C+\beta$, where however the constants $\alpha$ and $\beta$
are different with respect to $\Lambda$CDM, i.e. $\alpha\ne1$, in
general. Hence, on using the expression of $C$ given in (\ref{eq:conservation-eq})
and the definitions (\ref{eq:varrho_f}), we find at early times that
\begin{equation}
\lim_{z\to\infty}\Omega_{f}=1-\frac{f}{C\,f_{,C}^{2}}=1-\frac{f_{0,{\rm eff}}+(1+a_{1})\,(C-C_{0})}{(1+a_{1})^{2}C}\approx1-\frac{1}{1+a_{1}}\neq0\,,\label{eq:OmfET}
\end{equation}
which means that $f$ has non-trivial contributions at early times.
In particular, this implies that $\Omega_{r}$ does not go to unity
at early times, in general. Then one may wonder if this is enough
to rule out at once the model. To see this is not the case, let us
study the behavior of the effective equation-of-state for the $f(C)$
component at early times, for instance during radiation domination
when $\Omega_{r}\gg\Omega_{m}$. A direct calculation shows that 
\begin{eqnarray}
\lim_{z\to\infty}w_{f}\equiv\frac{p_{f}}{\varrho_{f}} & = & \frac{-\left[1-f_{,C}-2\,Cf_{,{\it CC}}\right]\sum_{i}(\varrho_{i}+p_{i})-\frac{1}{6}\,(f-C\,f_{,C}^{2})}{\frac{1}{6}\,(f-C\,f_{,C}^{2})}\nonumber \\
 & \approx & -1+\frac{a_{1}\sum_{i}(\varrho_{i}+p_{i})}{\frac{1}{6}\,[f_{0,{\rm eff}}+(1+a_{1})\,(C-C_{0})-(1+a_{1})^{2}\,C]}\nonumber \\
 & = & -1+\frac{6a_{1}\sum_{i}(\varrho_{i}+p_{i})}{f_{0,{\rm eff}}-(1+a_{1})\,C_{0}+(1+a_{1})C(1-1-a_{1})}\nonumber \\
 & \approx & -1+\frac{6a_{1}\sum_{i}(\varrho_{i}+p_{i})}{-(1+a_{1})a_{1}\,C}\approx-1-\frac{6H^{2}(1+w_{r})\Omega_{r}}{(1+a_{1})\,C}\nonumber \\
 & = & -1-\frac{-f_{,C}^{2}C(1+w_{r})\Omega_{r}}{(1+a_{1})\,C}=-1+(1+a_{1})(1+w_{r})\Omega_{r}\nonumber \\
 & \approx & -1+(1+a_{1})(1+w_{r})(1-\Omega_{f})=w_{r}=\frac{1}{3}\,,\label{eq:w_f}
\end{eqnarray}
where we have used Eq.\ (\ref{eq:OmfET}). Therefore this model gives
an effective contribution to radiation at early times. Therefore when
$w_{f}\approx1/3$, then $\varrho_{f}$ will contribute to the effective
radiation energy density $\varrho_{r}^{\text{eff}}=\varrho_{r}+\varrho_{f}$
so that, on the background, the effective $\Omega_{r}^{\text{eff}}$
will have an extra component due to $\Omega_{f}$ and $\lim_{z\to\infty}\Omega_{r}^{\text{eff}}=1$,
as we expect. We want to make clear that this model is not a model
of ``dark radiation'' for at least two reasons: 1) it gives an effective
radiation component only during radiation domination, whereas at late
times, $\rho_{f}$, on the background, behaves instead as an effective
cosmological constant; 2) no extra (massive or massless) degree of
freedom is introduced in the theory at any time, even during radiation
domination.

\section{Results}

\label{sec:results}

In this section, we present the results obtained from running a Monte
Carlo sampling of the parameter space for the kink model. 
Sampling the parameter space helps understanding the features
of the model, as otherwise we would find it hard to predict which
part of the parameter space is more appealing with respect to cosmology.
The fitness
parameter $\chi^{2}$ is an important quantity, which tells how much
the data prefer the model under investigation in the parameter estimation.
 Whenever a model has a better, i.e.~lower, $\chi^{2}$ than the standard cosmological
model $\Lambda\text{CDM}$, that model deserves attention in the context
of cosmological tensions. As we already wrote in the previous section,
we will see that the kink model \eqref{eq:asymmetricbump} consistently
gives a better fitness parameter than $\Lambda$CDM. For this reason,
we report on the kink model.

We present here our study of the behavior of both $\Lambda$CDM and
the kink model for several data sets constraining the evolution of
the background and perturbations both at high and low redshifts. We
used a Monte-Carlo sampling in order to find the bestfit parameter
points for both models which minimize the $\chi^{2}$. We gave very
broad priors on the free parameters of the kink model. In particular
we set $-1<a_{1}<50$ (with a flat prior), $0.5<\log_{10}a_{2}<18$
(with log-flat prior), and $-9<\log_{10}\beta<-0.75$ (with log-flat
prior) where $a_{3}=\beta\,a_{2}$. We set such large priors as the
model was to be applied to a large redshift range, without a priori
knowing whether the procedure would find any good fit at all in the
allocated sampling time. However, in the limit $a_{1}\to0$ the kink model reduces to
$\Lambda$CDM. We knew therefore that there should have been at least
some non-zero good-fit interval for the parameters. Despite those very 
broad priors, the procedure successfully converged to a good fit.

\subsection{Comparison of $\chi^2$}

We found that the kink model performs better than $\Lambda$CDM:
the fitness parameters $\chi^{2}$ (total and respective to each experiment)
are compared to those obtained in the $\Lambda$CDM case in Table
\ref{tab:all_data_bestfit}. 
\begin{table}
\begin{tabular}{|l|c|c|}
\multicolumn{1}{l}{Data sets $\downarrow$} & \multicolumn{1}{c}{$\chi^{2}$ for bestfit of $\Lambda$CDM} & \multicolumn{1}{c}{$\chi^{2}$ for bestfit of kink model}\tabularnewline
\hline 
Planck highl TTTEEE  & 2351.98  & 2339.45\tabularnewline
\hline 
Planck lowl EE  & 396.74  & 395.73\tabularnewline
\hline 
Planck lowl TT  & 22.39  & 20.84\tabularnewline
\hline 
JLA  & 683.07  & 682.98\tabularnewline
\hline 
bao boss dr12  & 3.65  & 3.66\tabularnewline
\hline 
bao smallz 2014  & 2.41  & 2.38\tabularnewline
\hline 
HST  & 13.03  & 11.63\tabularnewline
\hline 
All chosen data sets:  & in total $\chi^{2}=3473.27$  & in total $\chi^{2}=3456.67$\tabularnewline
\hline 
\end{tabular}\caption{Comparison of the total $\chi^2$ for the bestfit for both $\Lambda$CDM
and the kink model for all the considered data sets.}
\label{tab:all_data_bestfit} 
\end{table}
The difference in $\chi^{2}$ between the two models is
$\Delta\chi^{2}=16.6$. Even though we have three more parameters, we
find that the kink model is preferable and fits the data much better
than $\Lambda$CDM. To be more precise in Table~\ref{tab:chi2_info_cri}
we give two different information criteria for model
comparison~\cite{Trotta:2008qt,Arevalo:2016epc}, namely, Akaike
Information Criterion (AIC), and its corrected version,
AIC-c\footnote{It is well known that the Bayesian Information
  Criterion (BIC) gives a different penalty (typically larger) for
  models with a larger number of parameters. However BIC might
  penalize too much those models which have parameters unconstrained
  by data, as the one discussed here (in fact, the $\chi^2$ for the
  kink model is insensible to the relevant, i.e.\ small-enough, values
  of $\beta$, as we will show later on). For this reason, in terms of
  model selection, the AICc method seems to behave better especially
  regarding CMB data, as it gives similar results to other information
  criteria (e.g.\ the Deviance IC) (see \cite{Liddle:2007fy} for more
  on this point). If one would in any case use the BIC method, one
  would probably need to reduce the effective number of free
  parameters for the kink model, taking into account the 
  degeneracy for $\beta$.}. Accordingly, we have
$\Delta{(\textrm{AIC})} \approx 10.6$, and
$\Delta{(\textrm{AIC-c})} \approx 10.57$. This shows that the data
sets prefer the kink model (we have considered here a sample size of
order 3000). In the future, if experiments are not mistaken, the
discrepancy between $\Lambda$CDM and the kink model might increase, as
error bars may shrink. Of course, as an alternative, it is still
possible that some yet unknown systematics are tilting the balance in
disfavor of $\Lambda$CDM, especially for early time data.

\begin{table}
\begin{tabular}{|l|c|c|}
\multicolumn{1}{l}{Information Criteria $\downarrow$} & \multicolumn{1}{c}{$\chi^{2}_{\Lambda \textrm{CDM}}$} & \multicolumn{1}{c}{$\chi^{2}_{\textrm{kink}}$ }\tabularnewline
\hline 
AIC  & 3485.27  & 3474.67\tabularnewline
\hline 
AIC-corrected  & 3485.30  & 3474.73\tabularnewline
\hline 
\end{tabular}\caption{Comparison of $\Lambda$CDM and the kink model according to the Akaike information criterion and its corrected version.}
\label{tab:chi2_info_cri} 
\end{table}

Before studying the bestfit of the model, we want to make a few statements
about this result. First of all we can see that the kink model, performs
better than $\Lambda$CDM on the HST data point, thus alleviating
the tension in measurements of the value of today's Hubble factor
to some degree. The model performs in a way very similar to $\Lambda$CDM
for the other late time data, consisting mainly of BAO and Supernova
Type Ia data (JLA). The fact that the model is similar to $\Lambda$CDM
on these last experiments is afterall maybe not a surprise because the kink
model was constructed as to reduce to $\Lambda$CDM at late times
(with possibly different values for  the background parameters,
like \ $H_{0}$ for instance).

However, there is an important difference at early time  when $f_{,C}\neq1$. In fact,
we find that the main contribution to a lower value of the $\chi^{2}$
takes place in the redshift interval, which goes from today up to
values of redshift sensitive to Planck results, and the kink
model performs consistently better than $\Lambda$CDM for any of the
Planck experiments: $\Delta\chi_{{\rm TTTEEE}}^{2}=12.53$, $\Delta\chi_{{\rm lowl\ EE}}^{2}=1.01$,
and $\Delta\chi_{{\rm lowl\ TT}}^{2}=1.55$. The reason for such an
improvement is due to both a modification of the background (an effective
radiation component, which at late times changes into a cosmological
constant contribution, and, as we will see, to a fast change of the
value of $H$ at intermediate redshift $z\simeq743$), and 
to a different dynamics for the perturbations (whenever $f_{,C}\neq1$).
With these considerations, it is now time to focus our attention to the constraints and the bestfit
of the model in order to understand its characteristics.

\subsection{Two dimensional likelihoods and bestfit}

\begin{figure}[ht]
\includegraphics[width=13cm]{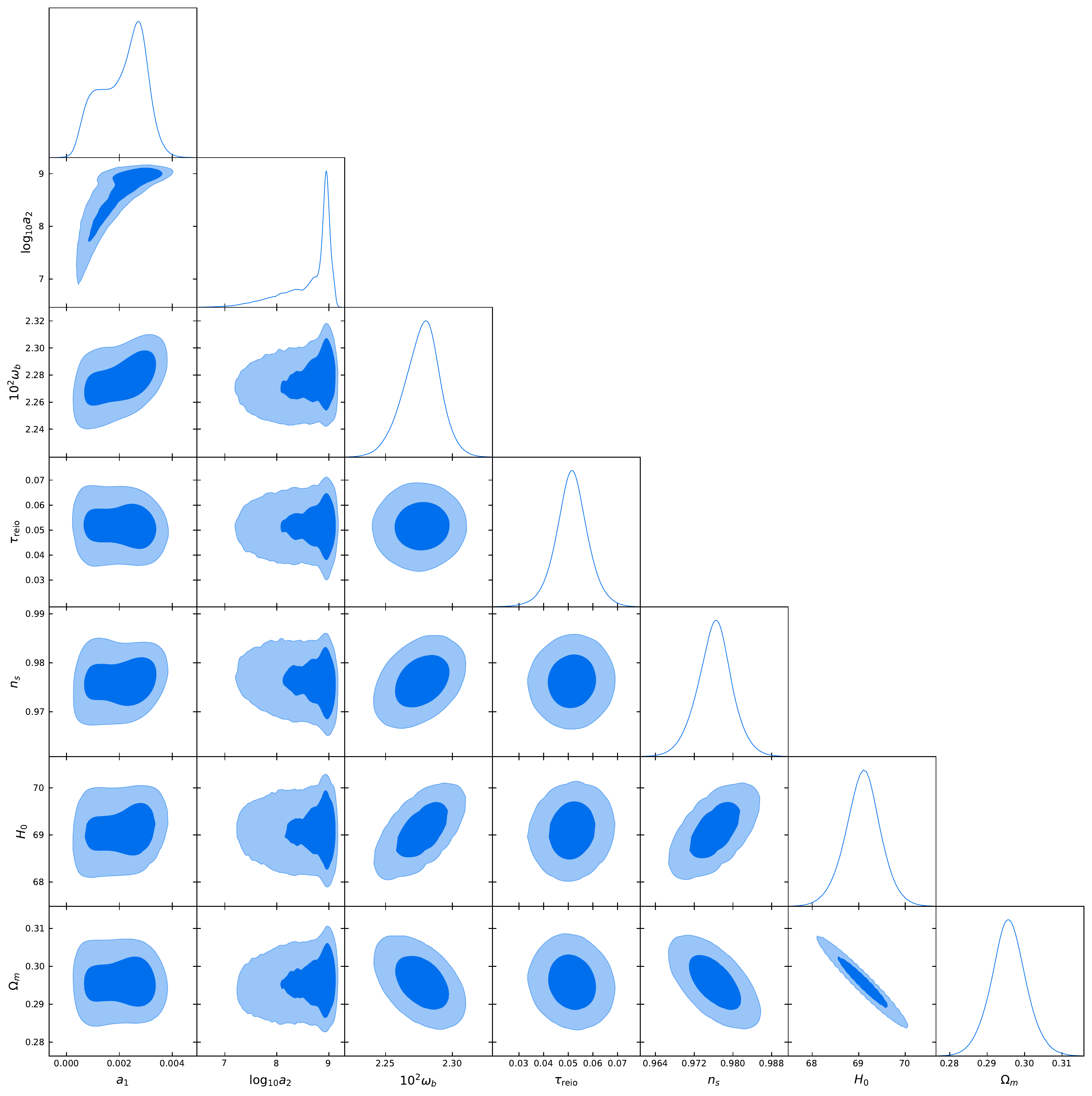} \caption{Marginalized two dimensional likelihoods for the kink model on fitting
all chosen data sets (both early and late time ones).\label{fig:Marginalized-all}}
\end{figure}

On analyzing the chains obtained after Monte-Carlo sampling we show
the two-dimensional marginalized likelihoods for the cosmological
variables of interest in Fig.\ \ref{fig:Marginalized-all}. 
The results for the one-dimensional 2$\sigma$ constraints are summarized
in Table \ref{tab:bestfit_alldata}. 
\begin{table}[ht]
\begin{tabular}{|l|c|}
\multicolumn{1}{l}{Parameters} & \multicolumn{1}{c}{95\% limits}\tabularnewline
\hline 
$a_{1}$  & $0.0028_{{-}0.0023}^{{+}0.0006}$ \tabularnewline
\hline 
$\log_{10}a_{2}$  & $8.95_{{-}1.33}^{{+}0.20}$ \tabularnewline
\hline 
$\log_{10}\beta$  & ${}<-3.5$ \tabularnewline
\hline 
$10^{2}\omega_{b}$  & $2.284_{{-}0.036}^{{+}0.019}$ \tabularnewline
\hline 
$\tau_{{\rm reio}}$  & $0.052_{{-}0.015}^{{+}0.013}$ \tabularnewline
\hline 
$n_{s}$  & $0.9778_{{-}0.0092}^{{+}0.0058}$ \tabularnewline
\hline 
$H_{0}$  & $69.19_{{-}0.90}^{{+}0.67}$ \tabularnewline
\hline 
$\Omega_{m}$  & $0.2952_{{-}0.0090}^{{+}0.0104}$ \tabularnewline
\hline 
\end{tabular}\caption{One-dimensional 2$\sigma$ constraints for the cosmological parameters
of interest obtained by fitting early times and late times data sets.}
\label{tab:bestfit_alldata} 
\end{table}

\begin{figure}[ht]
\includegraphics[width=5cm]{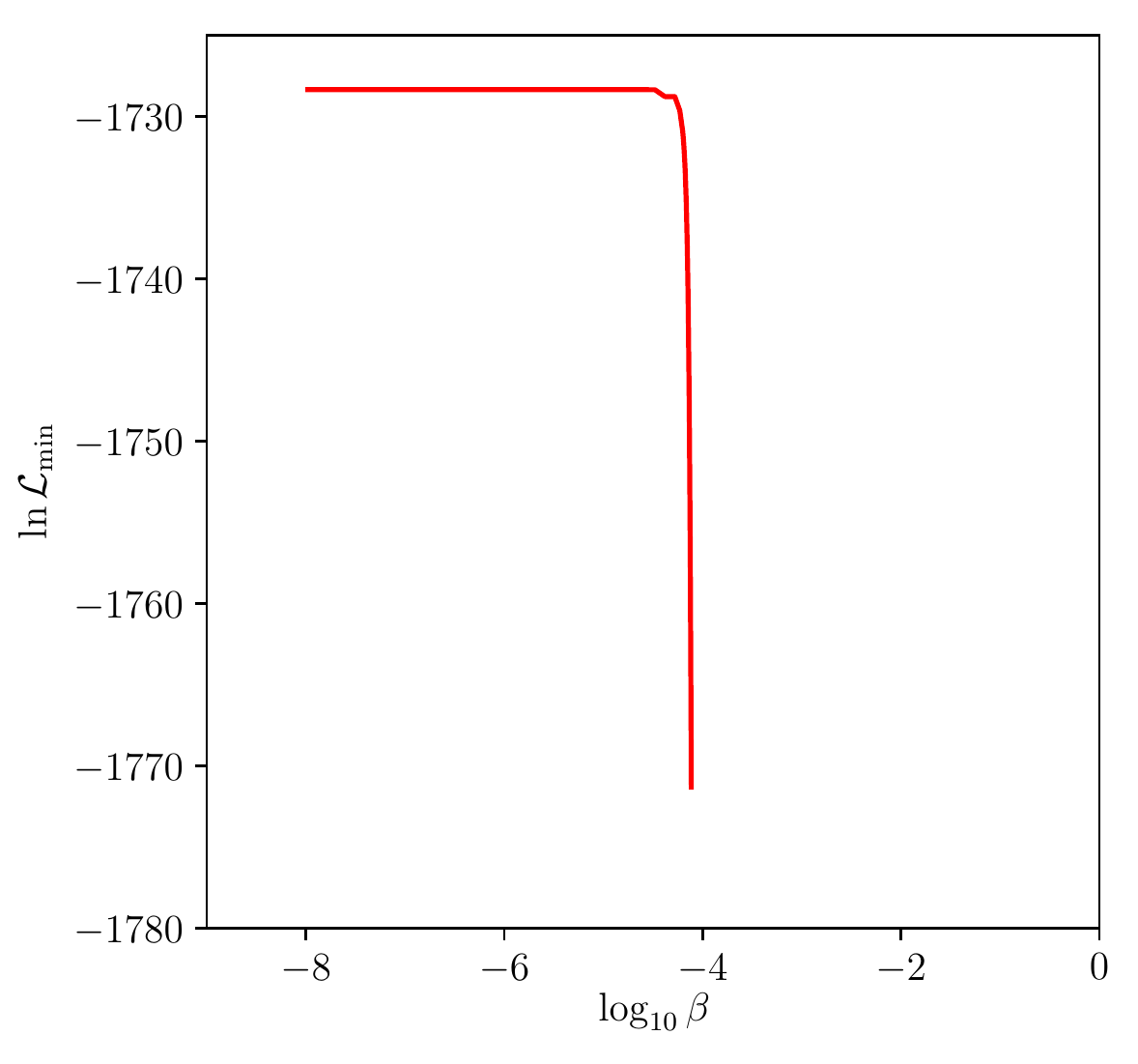} \caption{Independence of $\ln\text{\ensuremath{\mathcal{L}}}_{\text{min}}$ on $\beta$.\label{fig:beta_ind}}
\end{figure}

First of all, it is interesting to see the bound on $a_{1}$, which
evidently states that $a_1$ is different from 0 at 2$\sigma$, i.e. 
the data suggests non-negligible deviation from $\Lambda$CDM. The bound 
on $a_{2}$ is also interesting as we have a lower and an upper limit. In
particular, the value of $a_{2}$ selects the energy at which $f_{,CC}$
changes significantly. We will comment later on at which redshift
this change occurs. Finally let us briefly discuss about the parameter
$\beta$, which has been defined above as $a_{3}=\beta\,a_{2}$. The value of $\beta$ does not affect
the value of the $\chi^{2}$ when it is sufficiently small. In fact, in order to study better the
dependence of the $\chi^{2}$ on the parameter $\beta$, we have performed
the experiment of changing the value of $\beta$ for the best fit
and confirmed that its value does not affect the minimum of $\chi^2$, as shown in Fig.~\ref{fig:beta_ind} which plots the behavior of the likelihood when $\beta$ varies, as long as $\log_{10}\beta\apprle-4.1$, for higher precision.

This means that we have a degeneracy for small values of $\beta$.
In fact, the allowed lower bound was hitting the lower prior limit
(which was set to be $10^{-9}$). Much smaller values would in fact require
a sufficiently high precision which is not allowed in our computations.
The meaning of a small $\beta$ is anyhow simple, the redshift range
for which $f_{,CC}$ changes significantly cannot be too large, i.e.\ the
transition between the two different $f_{,C}$ cannot be adiabatic.
Later on, when we study in detail the minimum for the whole data sets,
we will also try to understand the reason why large values of $\beta$ seem to
be excluded.

As already stated above for these all-redshift-range data sets, we
can see that the bestfit kink model alleviates the $H_{0}$ tension
slightly, but, mostly, it gives a better fit on Planck data compared
with $\Lambda$CDM. In fact, a value for $\Delta\chi^{2}=16.6$ is
large enough (even on considering three more parameters for the kink
model) as to exclude $\Lambda$CDM at 2$\sigma$ (because $a_{1}$
is different from 0 at 2$\sigma$, and the $a_{1}\to0$ limit is a
smooth one, as we have no other propagating mode than $\Lambda$CDM). This result is encouraging for the kink model and at least tells
that flat-$\Lambda$CDM does not necessarily give the better
fit to Planck data, or we may conclude that, if flat-$\Lambda$CDM
is the only allowed model a priori, Planck data indicate some unknown
systematics. A simple solution, in this sense following Occam's razor,
is that models different from $\Lambda$CDM have an arena to test
gravity already in the Planck data sets and at high redshifts.

In summary, we have found a model which fits Planck data better than
flat $\Lambda$CDM. The fact that there are models which fit Planck
data better than flat $\Lambda$CDM is not new (see e.g.\
\cite{DiValentino:2019qzk,DiValentino:2020hov,Peirone:2019aua}, the
last one describing a model in the context of Horndeski theories). For
example, it is now well known that Planck data prefer
$\Omega_{K}\neq0$ \cite{Aghanim:2018eyx,DiValentino:2019qzk}. However,
a non-flat model is then in strong tension with BAO and inflationary
paradigm at the same time \cite{DiValentino:2020hov}. On the contrary, we have shown here that
BAO data do not constrain much this model (or they constrain it as
much as flat-$\Lambda$CDM) and, as a consequence, do not spoil the
fact that the kink model still performs better than flat $\Lambda$CDM.

\subsection{Background evolution}

\begin{figure}[ht]
\includegraphics[width=13cm]{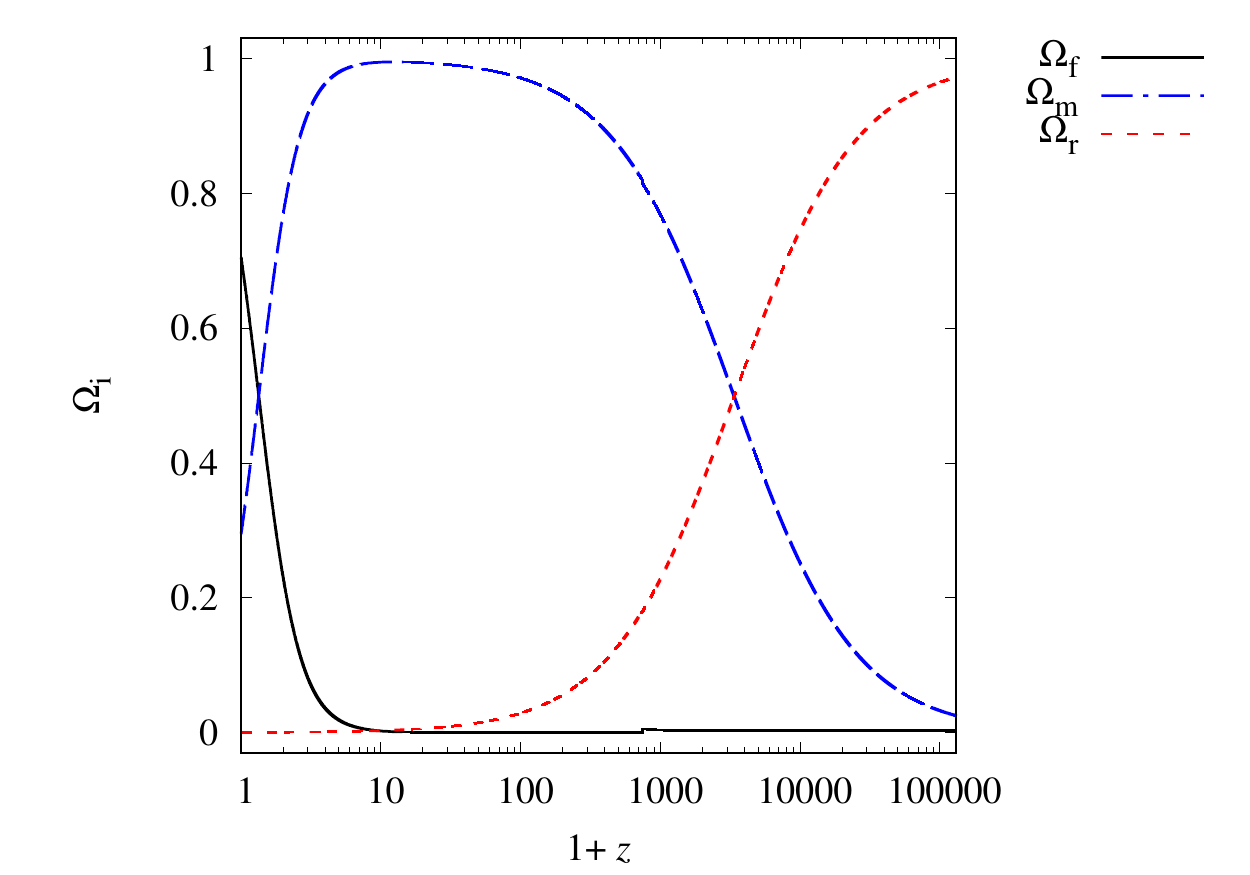} \caption{Evolution of the several $\Omega_{i}$ ($i=f,m,r$) functions during
the evolution of the background for the bestfit kink model. Around
$z\approx743$ some non trivial behavior can be seen. Such a behavior
will be studied more in detail in the following. The plot has been
obtained on fixing $h=0.691867$, $\Omega_{m0}=0.2951803$, $a_{1}=2.766667\times10^{-3}$,
$\log_{10}a_{2}=8.949697$, $\log_{10}\beta=-4.5$, $\Omega_{f0}=0.7047323$.\label{fig:BF_Omegas}}
\end{figure}

In the following we study in detail the behavior of the minimum of
the $\chi^{2}$ starting with the background evolution. We have fixed
the various background parameters to their bestfit values and integrated
the background equations of motion. The results of the evolution
of the three $\Omega_{i}$ functions are given in Fig.\ \ref{fig:BF_Omegas}. 

\begin{figure}[ht]
\subfloat[Evolution of $\epsilon_{H}$]{\includegraphics[width=8cm]{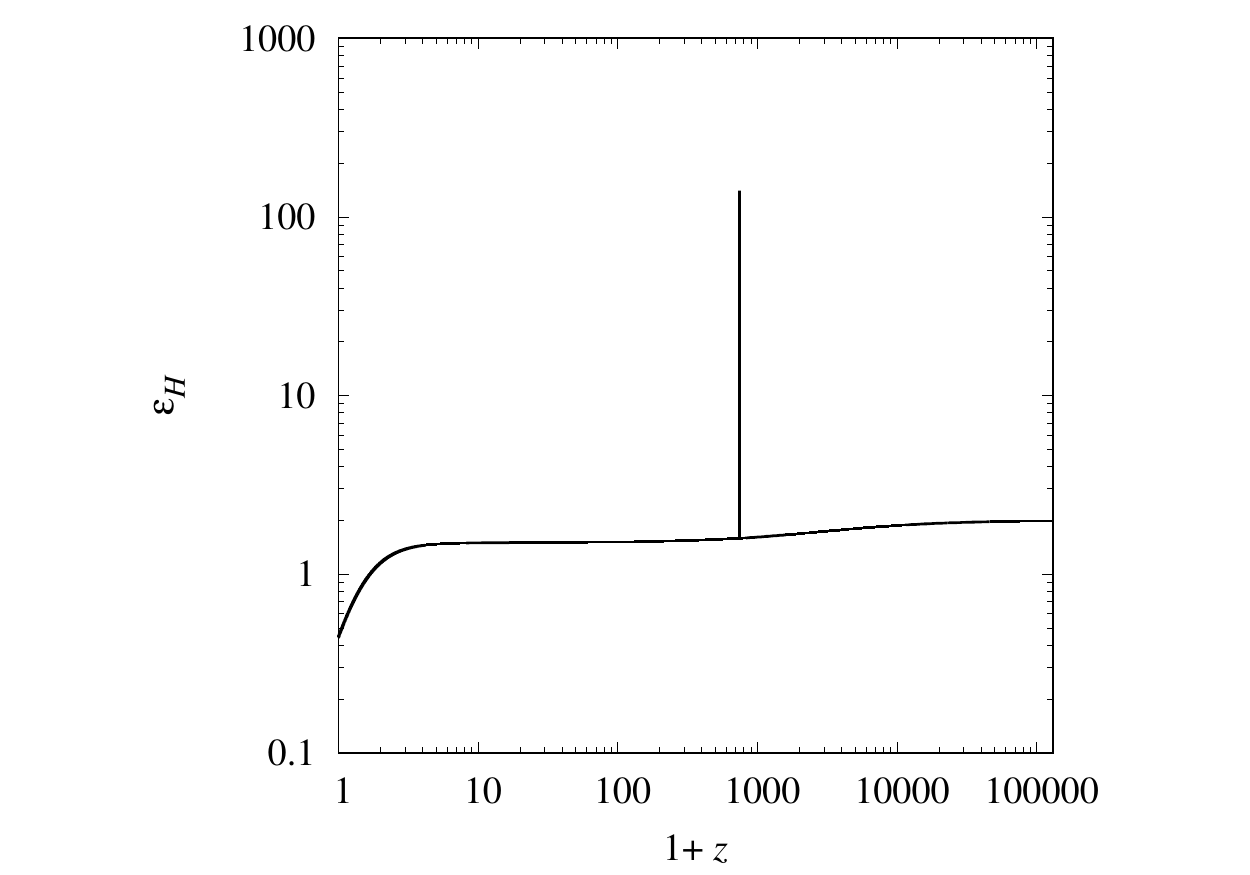}

}\subfloat[Evolution of $\epsilon_{H}$ (zoomed version)]{\includegraphics[width=8cm]{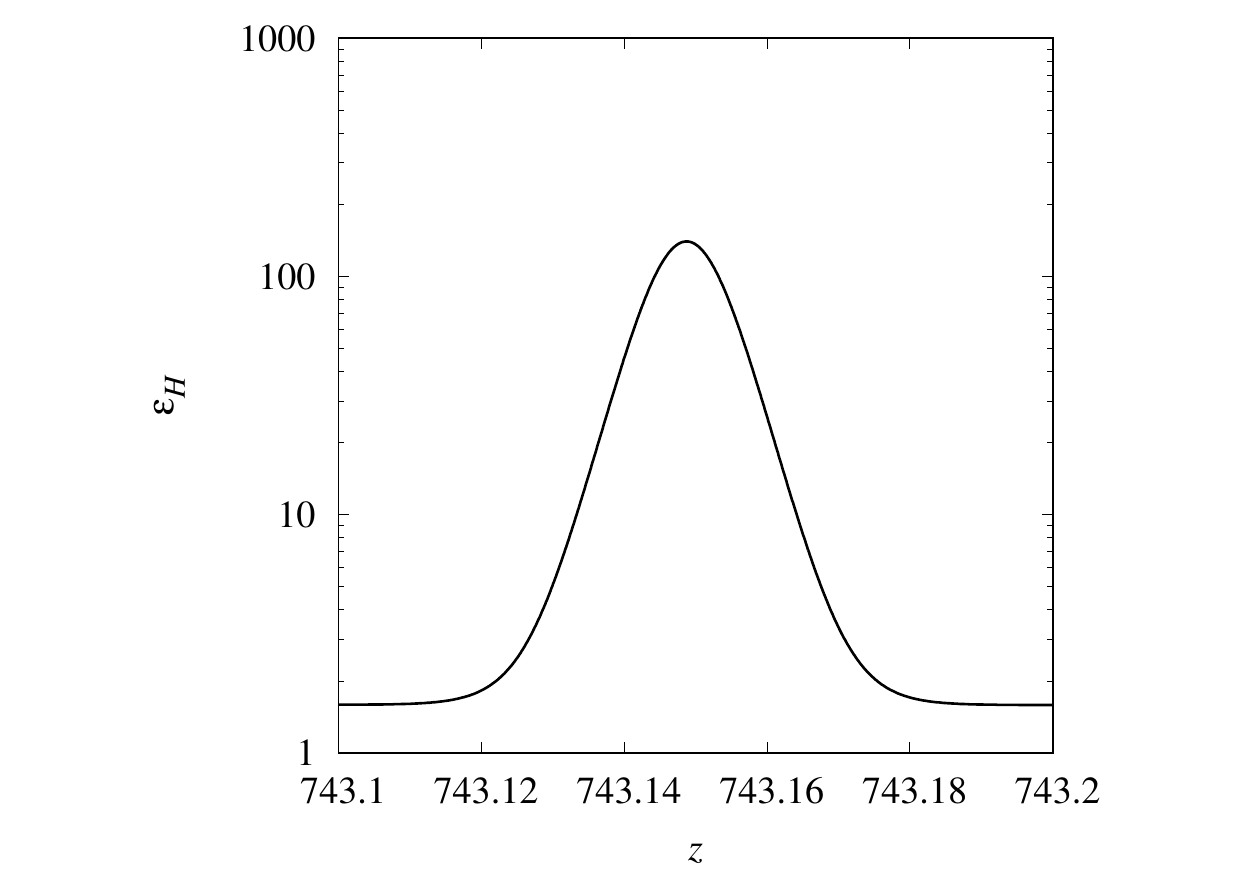}

}\caption{Dynamics of the background variable $\varepsilon_{H}=-\dot{H}/(NH^{2})$.
On the left panel we can see that today the universe accelerates (since
$\varepsilon_{H}<1$), whereas at early times the universe is radiation
dominated (as $\varepsilon_{H}\to2$). On the right panel, we zoom
around the redshift $z\simeq743$, and we show that the transition
is fast but smooth, i.e.\ no singularity is present (and this also
implies that numerics are stable). \label{fig:eps_H}}
\end{figure}

We can see that some non-trivial dynamics happens at intermediate
redshifts ($z\simeq743$). In order to understand better what happens
we further study the dynamics of the background for several variables.
One variable of interest is $\varepsilon_{H}\equiv-\dot{H}/(NH^{2})$
(which can be calculated via Eq.\ (\ref{eq:secondaE})), as it shows
whether the universe accelerates (when $\varepsilon_{H}<1$), decelerates
(approaching 2 during radiation domination), and if some non-trivial
singularity is present (besides the big-bang). In Fig.\ \ref{fig:eps_H},
we can see that around $z\simeq743$, there is a fast but smooth transition
at which $\varepsilon_{H}$ grows but remains finite. Therefore no
singularity is present. Furthermore, we can see that this behavior
takes place in an interval $\Delta z<1$. Finally this variable shows that the universe
starts being radiation dominated, then matter dominated, and finally
the universe starts accelerating. 

\begin{figure}[ht]
\subfloat[Evolution of $w_{f}$]{\includegraphics[width=8cm]{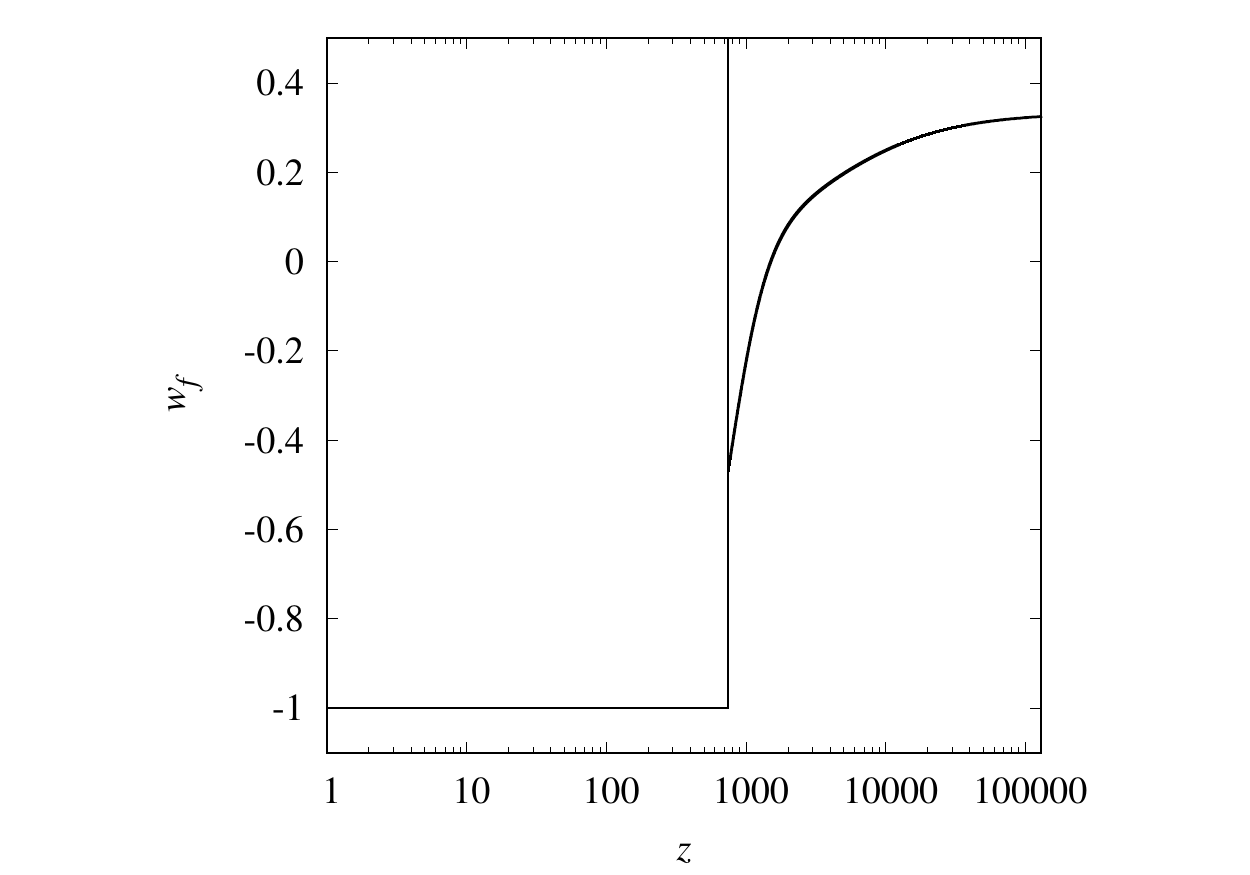}

}\subfloat[Evolution of $w_{f}$ (zoomed version)]{\includegraphics[width=8cm]{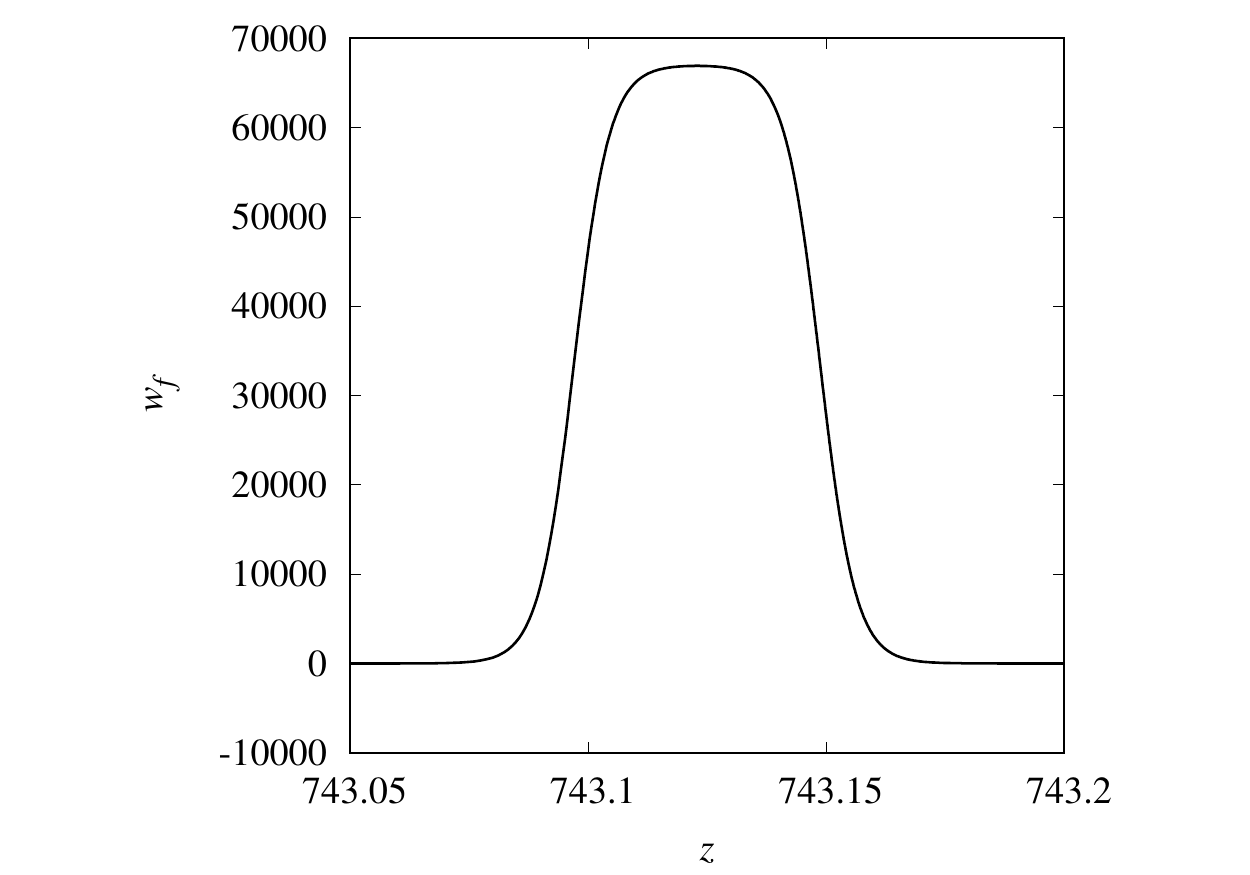}

}
\caption{Dynamics of the equation of state variable $w_{f}$ for the $f$-component.
On the left panel we can see that today the universe tends to be dominated
by a cosmological constant (since $w_{f}\approx-1$), whereas at early
times the $f$-component tends to behave as a radiation-component
as $w_{f}\approx1/3$. On the right panel, we zoom around the redshift
$z\simeq743$, and we show that also for $w_{f}$ the transition is
fast and smooth. \label{fig:w_f}}
\end{figure}

Another variable of interest is represented by $w_{f}=p_{f}/\rho_{f}$,
i.e.\ the effective equation of state for the $f$-component. Its
behavior is shown in Fig.\ \ref{fig:w_f}, where we can see that
the $f$-component, as already mentioned before, behaves as radiation
at early times, but as a cosmological constant at late times. 

\subsection{Evolution of the perturbations}
\begin{figure}[ht]
\subfloat[Evolution of $c_{s}^{2}$ around the transition point]{\includegraphics[width=8cm]{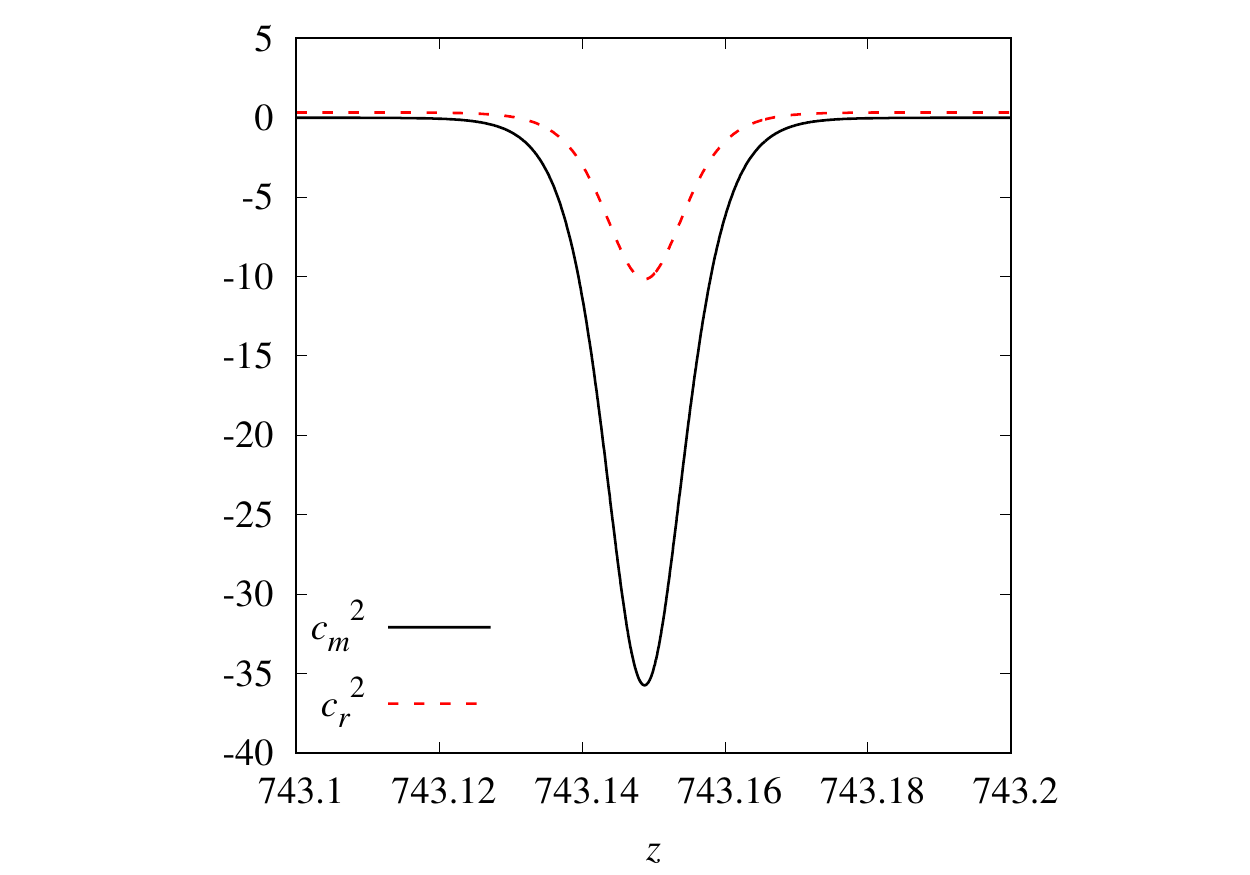}

}\subfloat[Evolution of $c_{T}^{2}$ around the transition point]{\includegraphics[width=8cm]{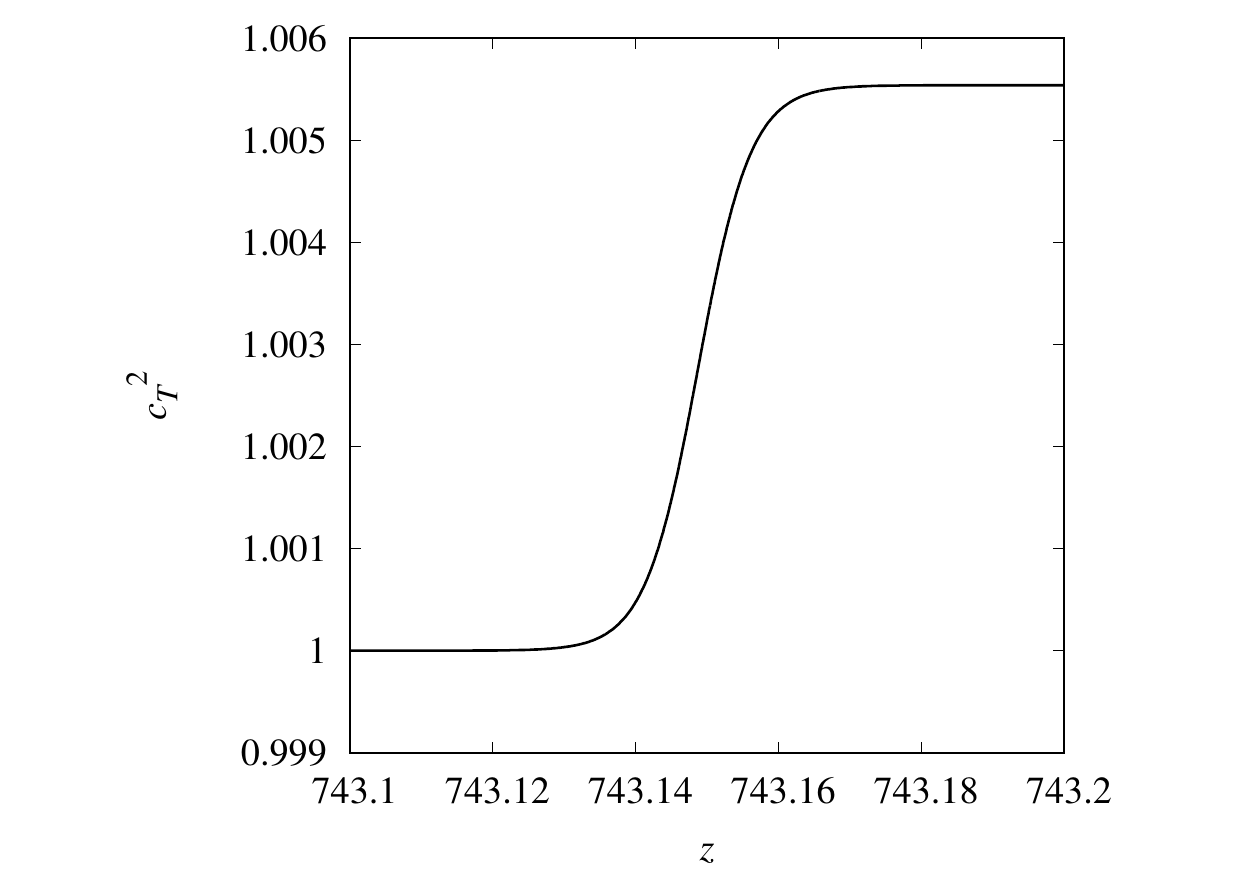}

}
\caption{Evolution of the squared speeds of propagation for the scalar-matter
modes and tensor modes. On the left panel we can see that the squared
speeds of propagation for both dust ($c_{m}^2$) and radiation ($c_{r}^2$) change, and their
values become negative during the transition. For a short time gradient
instabilities take place, but as we will see later on, it is not so
catastrophic (because the transition is short enough) as to make the
matter perturbations totally unstable. This implies that $\beta$
(the parameter which determines how fast the transition is) must be
small enough so that the instability-era is negligible. On the right
panel, we show instead that at early times the speed of tensor modes
was not unity, but today it is indistinguishable from unity as to
pass the multi-messenger constraints. \label{fig:c2_sT}}
\end{figure}
\begin{figure}[ht]
\subfloat[Evolution of $\delta_{\gamma}$]{\includegraphics[width=8cm]{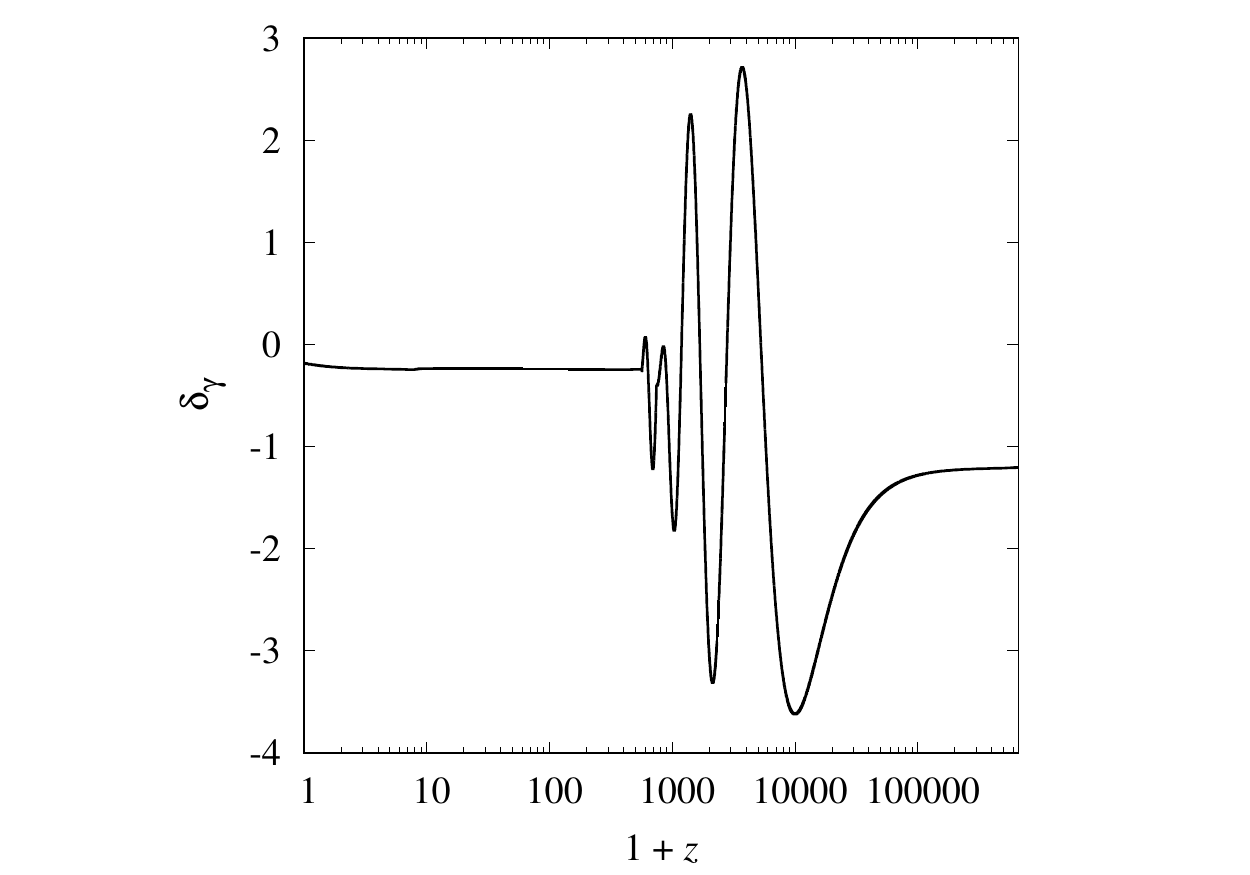}

}\subfloat[Evolution of $\delta_{\gamma}$ around the transition point]{\includegraphics[width=8cm]{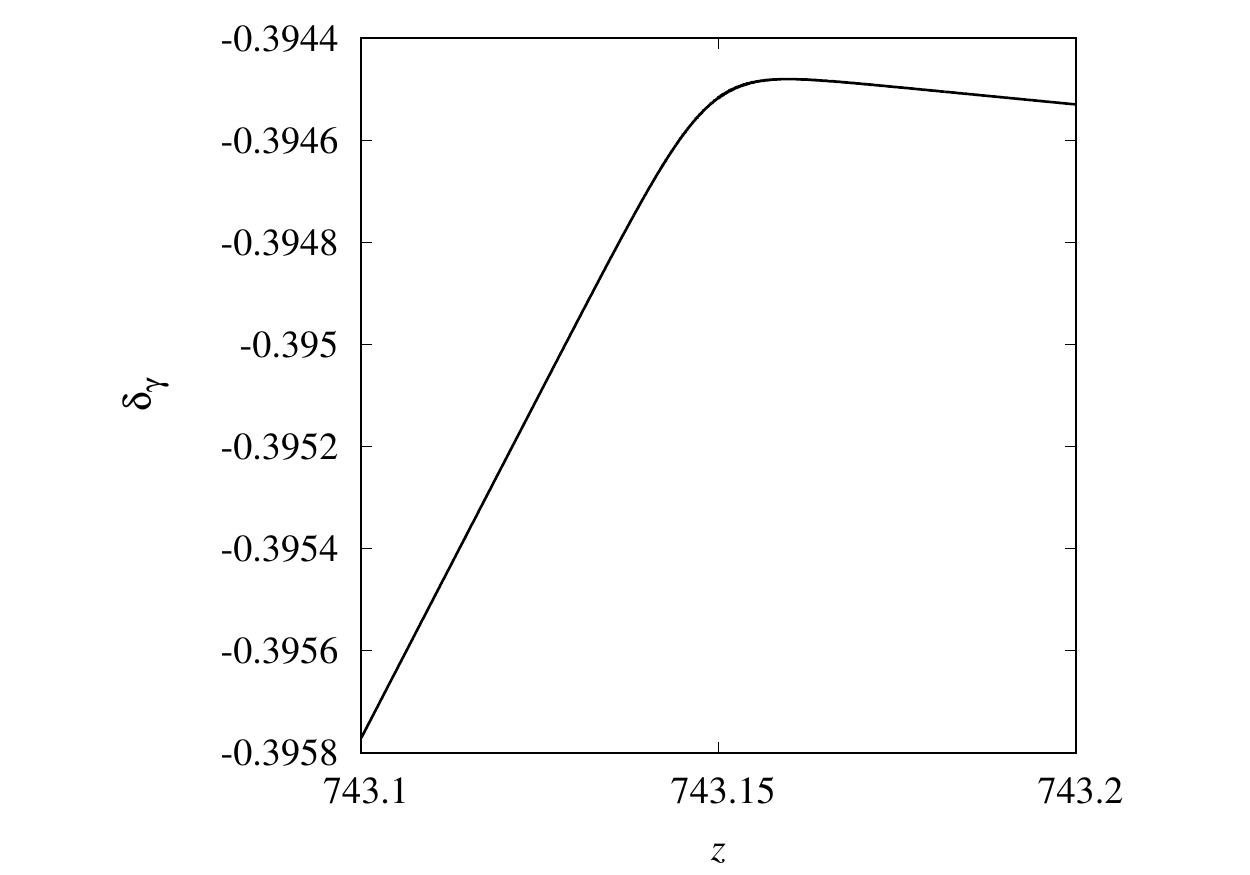}

}
\caption{Dynamics of the perturbation variable $\delta_{\gamma}$, the photon
energy density contrast. It is clear from both panels (the right one
describing its evolution during the transition-era), that such a variable
still evolves in a way similar to GR after this transition ends).
For this plot we have used the same bestfit parameters of the background,
and on top of that $\omega_{b}=0.02284032$, $\omega_{c}=0.1184566$,
$\tau_{{\rm reio}}=0.05183895$, $\ln(10^{10}A_{s})=3.039270$, $n_{s}=0.9778259$,
and, as for the size of the wave-vector, $k=0.1$ Mpc$^{-1}$ (to
enhance the effect of the instability in the very short wavelengths,
but still linear).\label{fig:deltaG}}
\end{figure}
\begin{figure}[ht]
\subfloat[Evolution of $\theta_{\gamma}$ and $\sigma_{\gamma}$]{\includegraphics[width=8cm]{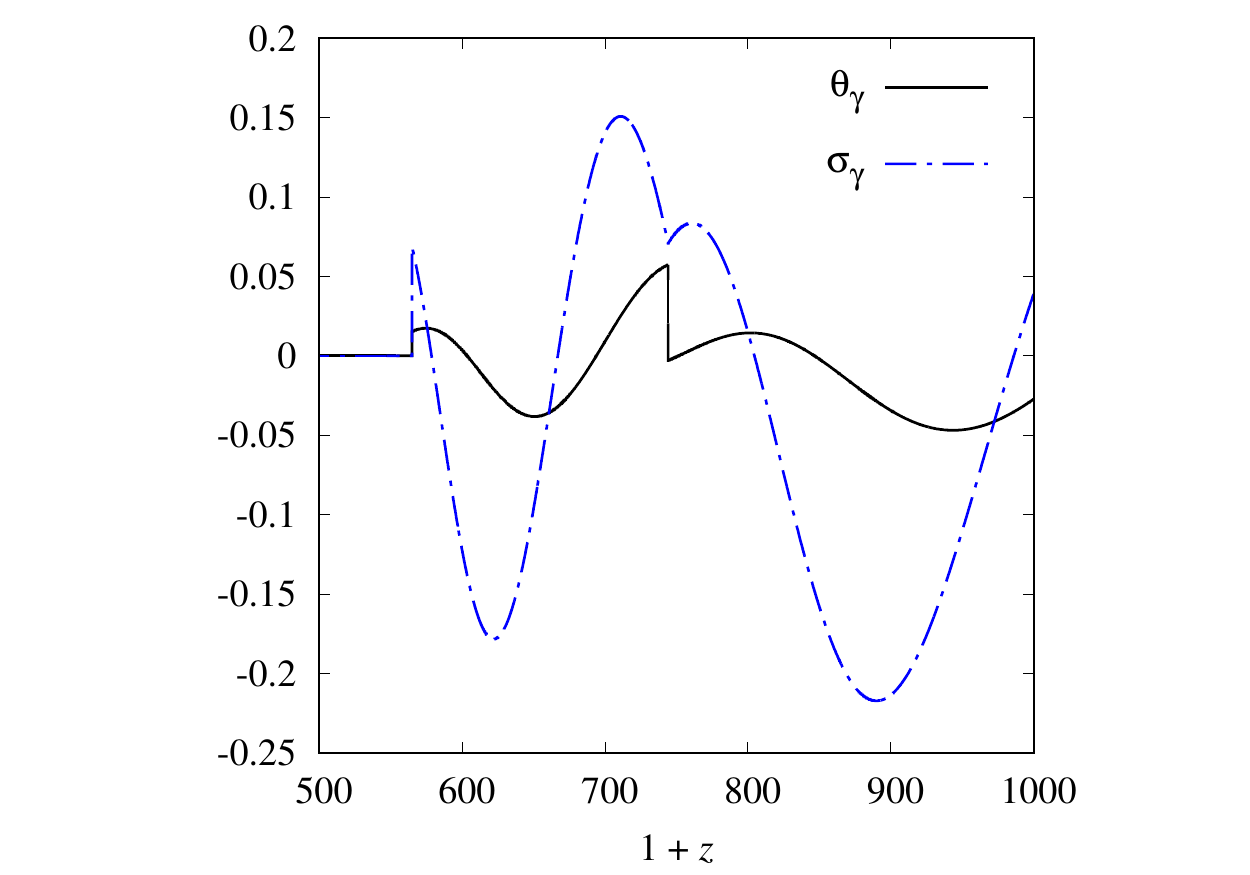}

}\subfloat[Evolution of $\theta_{\gamma}$ and $\sigma_{\gamma}$ around the
transition point]{\includegraphics[width=8cm]{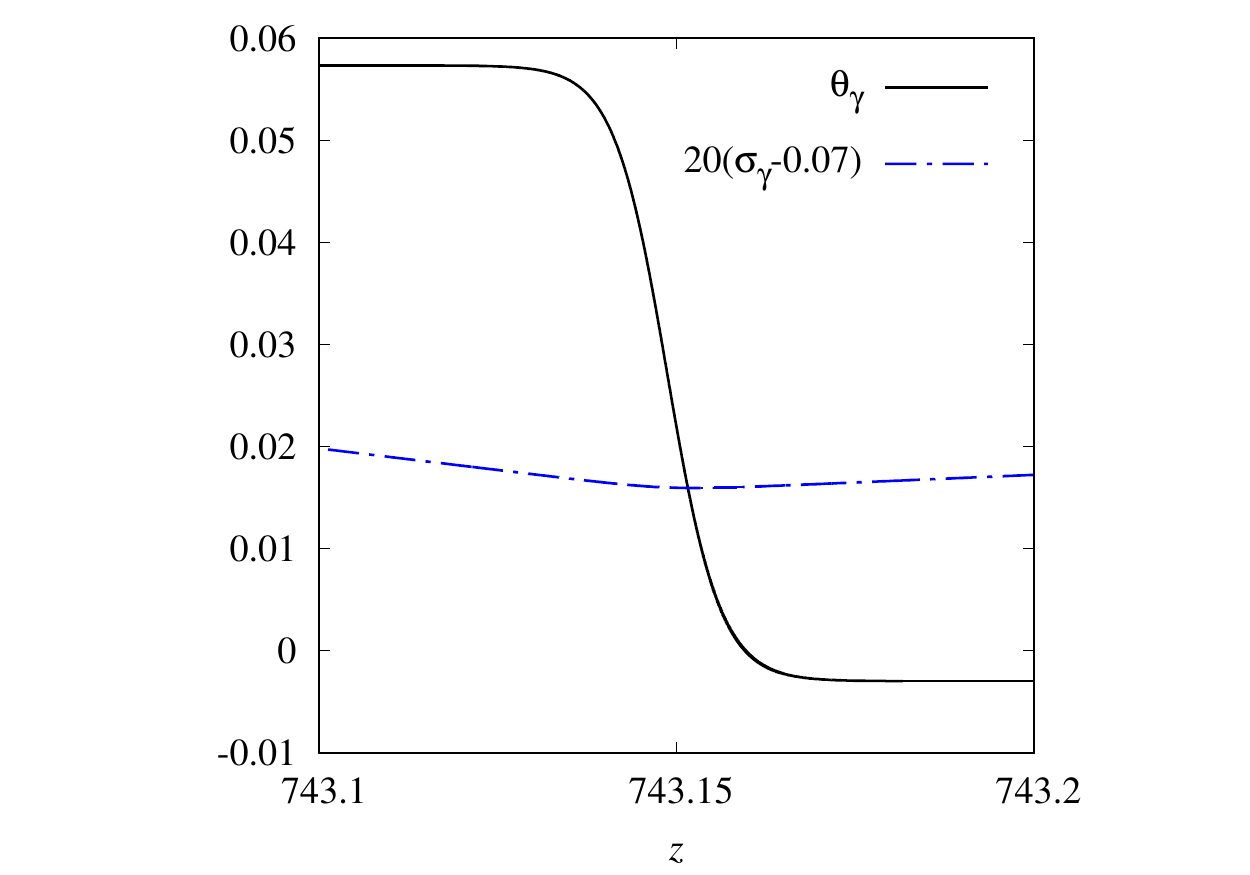}

}\caption{Dynamics of the perturbation variable $\theta_{\gamma}$, related
to the photon velocity perturbation, and the photon shear $\sigma_{\gamma}$.
For low redshifts, these variables become more and more negligible,
as expected from their behaviors in GR. In the right panel, in order
to enhance the change in the shear (with respect to the change of
$\theta_{\gamma}$) we have multiplied its difference about the mean
value at the transition era ($\approx0.07$), by a factor of 20.\label{fig:thetaG}}
\end{figure}

\begin{figure}[ht]
\subfloat[Evolution of $\theta_{b}$ and $\theta_{c}$]{\includegraphics[width=8cm]{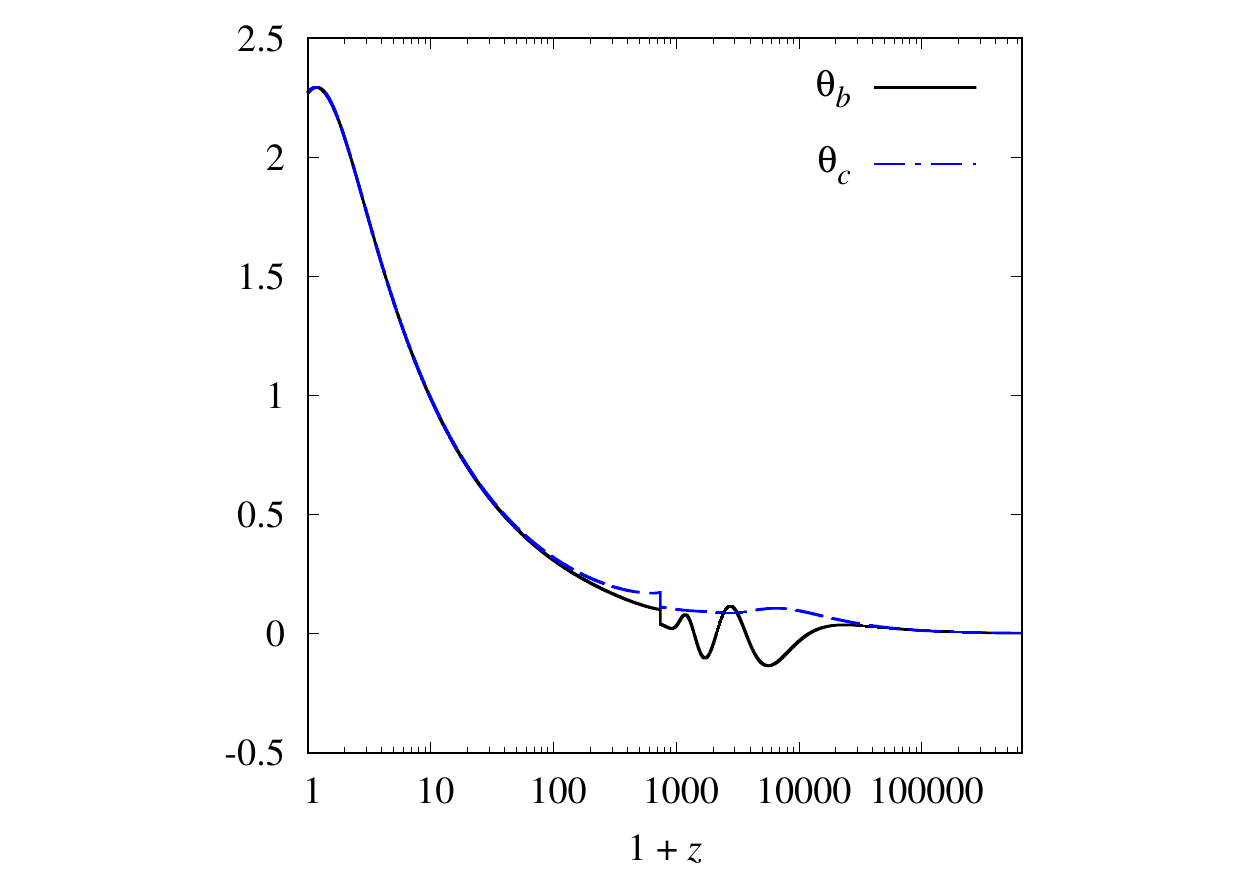}

}\subfloat[Evolution of $\theta_{b}$ and $\theta_{c}$ around the transition
point]{\includegraphics[width=8cm]{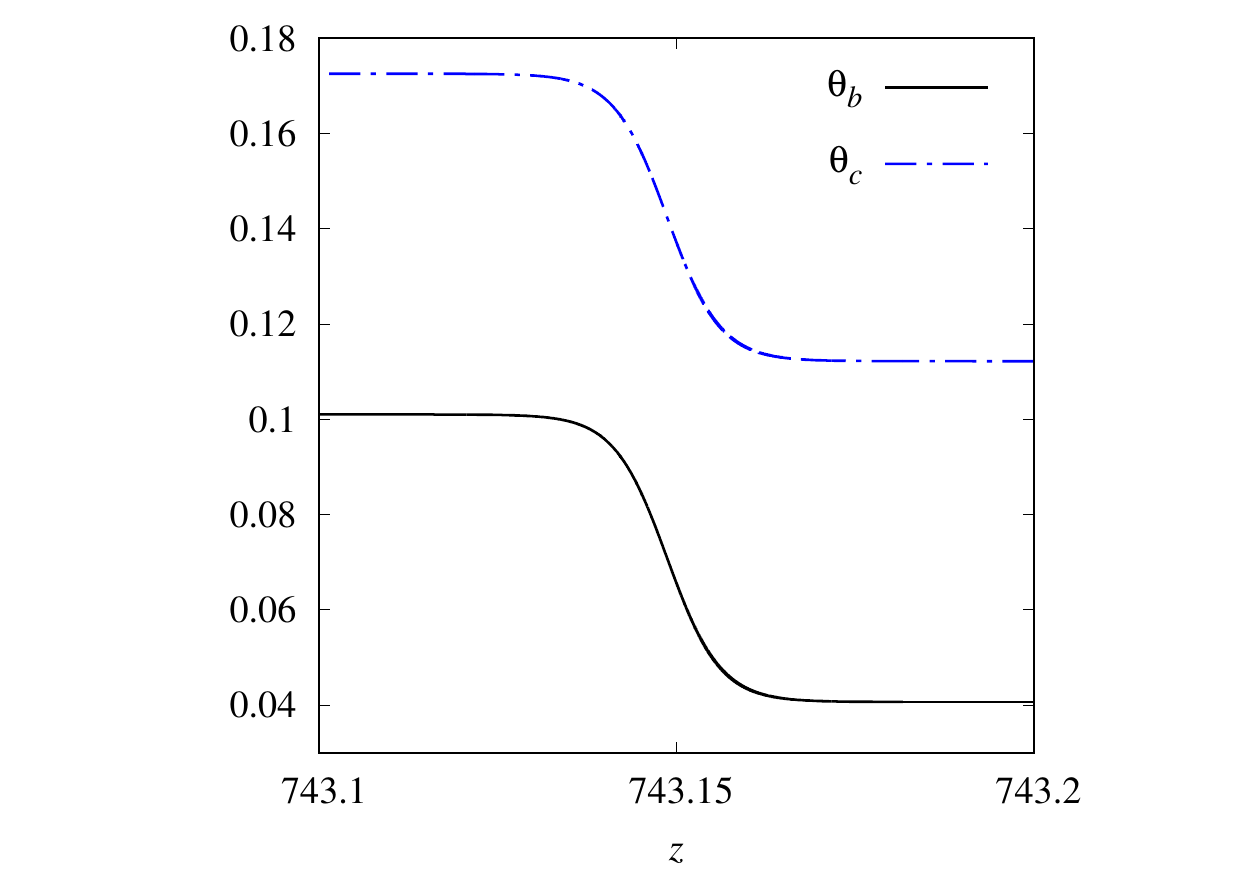}

}
\caption{Dynamics of the perturbation variables $\theta_{b}$ and $\theta_{c}$
(related to the velocity perturbations for the baryons and the CDM
components respectively). Also for these variables, after the transition-era
ends, their evolution is not too different from GR.\label{fig:thetabc}}
\end{figure}

\begin{figure}[ht]
\subfloat[Evolution of $\delta_{b}$ and $\delta_{c}$]{\includegraphics[width=8cm]{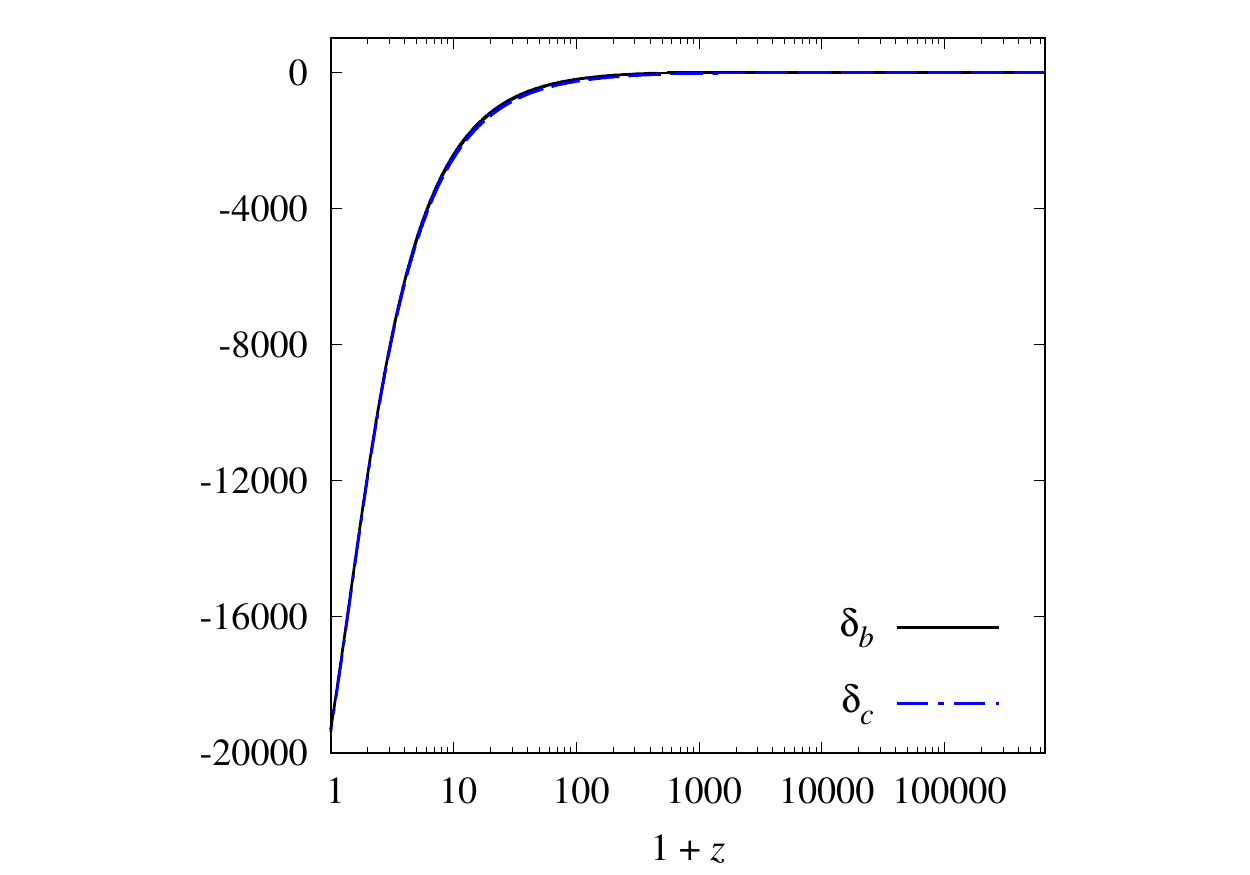}

}\subfloat[TT correlation dimensionless coefficients $C_{l}^{{\rm TT}}$, where
any sudden transition has completely disappeared]{\includegraphics[width=8cm]{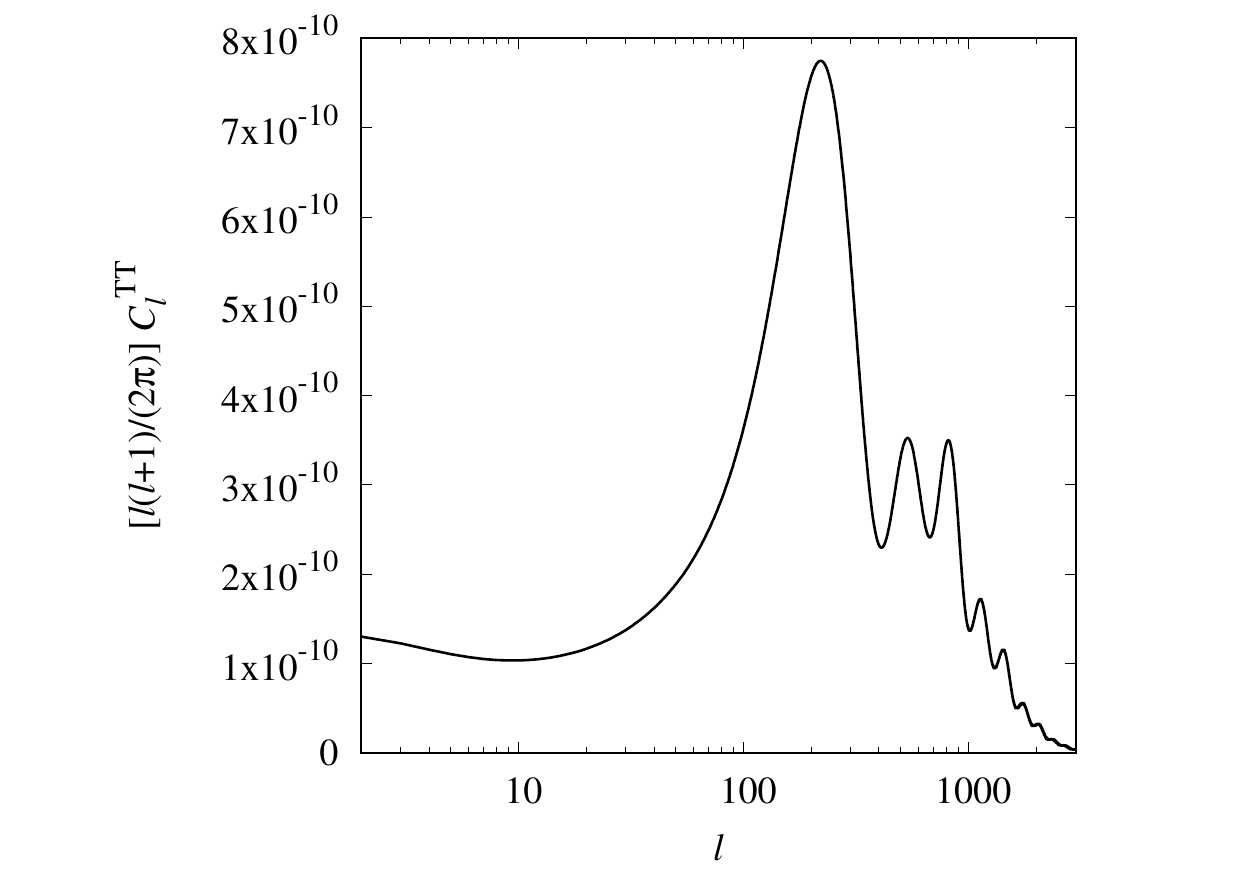}

}
\caption{Left panel: dynamics of the perturbation variables $\delta_{b}$ and
$\delta_{c}$ (i.e.\ the density contrasts for baryons and CDM).
Their evolution is dominated by the Jeans instability which leads
to galaxy formation, in a way still consistent with GR. Right panel:
TT correlation dimensionless coefficients $C_{l}^{{\rm TT}}$ which
are bound by observations. This variable, involving typically an integral
over the line of sight, smooths out the transition era for the perturbations.\label{fig:Cls}}
\end{figure}

After having studied in details the behavior of the background, we now focus on
the dynamics of the perturbations. First of all, let us see
what happens for the speed of propagation for the modes. In Fig.\ \ref{fig:c2_sT},
we can see that the perturbation variables undergo an era of instability
(because $c_{s}^{2}<0$), whereas tensor modes acquire a velocity
of propagation different from unity (at early times, i.e.\ for $z>743$).
The fact that there is an era of instability should make us worried,
however such an era is relatively short\footnote{We find that in cosmic time, for the bestfit, this era lasts $\Delta t\simeq\frac{|\Delta z|}{1+z}\,\frac{H_{0}}{H}\,H_{0}^{-1}$.
For $H_{0}^{-1}=9.7776\times10^{9}\,h^{-1}$ yr, and $|\Delta z|\simeq0.1$,
we find $\Delta t\approx155.7$ yr.}. Therefore, there is the chance that the matter perturbations still
remain finite after that transition-era ends. This is exactly what
happens. After all, we already know that the $\chi^{2}$ is really
good for this model. In fact, the $\chi^{2}$ from Planck comes from
correlations of quantities (for example TT correlations) which typically
involve an integral over the line-of-sight. Therefore, we expect that,
if perturbations follow again GR evolution after the transition-era,
the results for any such integral will be smooth. Since linear perturbations
remains under control, one may expect a similar behavior for higher
order perturbations, although we have not studied here this complicated
issue. 

Let us then study the evolution of the perturbation variables. As
shown in Fig.\ \ref{fig:deltaG}, the photon energy density contrast
$\delta_{\gamma}$ remains finite and once the transition is over,
its evolution returns to the one similar to GR. 
Similar considerations can be made for the other perturbation variables,
as shown in Figs.\ \ref{fig:thetaG}, \ref{fig:thetabc} and \ref{fig:Cls}.

To conclude the study of the best fit model, we want to emphasize
again that the transition era is short-lived enough as to make the
instability inefficient so that as soon as it ends, the dynamics is
once again following the GR evolution. Therefore this model changes
the $\chi^{2}$ not because of the transition, but because of the
different behavior which it shows at early times (i.e.\ for $z>743$).
It should be noticed that the numerics have been always stable as
to be sensitive to small variations of $a_{1}$ (which on the bestfit
is about ${+2}\times10^{-3}$, whereas $a_{1}=0$ corresponds to $\Lambda$CDM)
and of the parameter $\beta$ governing the duration of the transition
era. This sensitivity has been achieved by reducing the standard tolerance
parameters ($\simeq10^{-5}$) which determine the precision of the
numerical solution of the Boltzmann solver, CLASS, by about ten orders
of magnitude.

\section{Conclusion }

\label{sec:conclusion}

We have studied a ``kink model'' within the class of minimally modified
gravity theories dubbed $f(\mathcal{H})$ theories, which was shown
in the present work to have a fit to several early times and late
times data sets better than $\Lambda$CDM by $\Delta\chi^{2}=16.6$.
We are impressed by several aspects of it. Most importantly, this
model shows once for all that Planck data and $\Lambda$CDM do not
fit so well with each other. In particular, the kink model performs
better than $\Lambda$CDM especially on Planck 2018, but also on $H_{0}$
single-point data sets.

In the context of $\Lambda$CDM this aspect is ameliorated for Planck
data alone notably if one introduces a non-zero curvature term, which
however is not well motivated by inflation and moreover it is strongly
suppressed by late time data (especially BAO). On the other hand,
late time data do not strongly constrain the kink model, at most as
much as flat-$\Lambda$CDM, and the kink model still keeps a better
fit to Planck 2018 data.

On top of this consideration, the bestfit kink model is characterized
by the presence of a transition era which happens at a redshift of
$z\simeq743$\footnote{Inside the 2$\sigma$ region for $a_{2}$ we find that the redshift
of transition era, $z_{{\rm TE}}$, occurs in the range $278<z_{{\rm TE}}<859$,
i.e.\ at intermediate redshifts.}. Such a transition era happens at intermediate redshifts and not
at very high energies. This means that there is the possibility that
such a transition may occur in other contexts with strong gravity
(inside a star for example). Furthermore during this transition era,
the scalar perturbations have an instability ($c_{s}^{2}<0$) which,
at least, as long as linear perturbation theory holds, does not seem
to be catastrophic because of its short duration ($\Delta z\simeq0.1$
around $z\simeq743$) so that the matter perturbations (for which
we have many constraints) do not show any divergent behavior. In particular,
this transition era seems to be washed out when we look at observables
which imply a line-of-sight integration.

The kink model then opens up an arena for modified gravity models,
which so far are introduced mostly to address late time data only.
This model in fact, addresses and resolves tension for $\Lambda$CDM
in Planck data and at the same time shows a non-trivial window for
phenomenology at intermediate redshifts. Future work is needed to
address several points which remain to be explored, such as the existence,
composition and evolution of compact objects for this model.

\begin{acknowledgments}
  The work of K.A. was supported in part by Grants-in-Aid from the Scientific Research Fund of the Japan Society for the Promotion of Science, No.~19J00895 and No.~20K14468.
  The work of A.D.F.\ was supported by Japan Society for the Promotion
  of Science Grants-in-Aid for Scientific Research No.\ 20K03969. The
  work of S.M. was supported in part by Japan Society for the
  Promotion of Science Grants-in-Aid for Scientific Research
  No.~17H02890, No.~17H06359, and by World Premier International
  Research Center Initiative, MEXT, Japan. KN acknowledges support from the CNRS project 80PRIME. M.O. would like to thank
  Ruth Durrer for her hospitality during the preparation of this work,
  and for her insightful comments. M.C.P. acknowledges the support
  from the Japanese Government (MEXT) scholarship for Research
  Student. Numerical computation in this work was carried out at the
  Yukawa Institute Computer Facility.
\end{acknowledgments}

 \bibliographystyle{unsrt}
\bibliography{bibliography}

\end{document}